\documentclass[aps,prd,preprint,nofootinbib]{revtex4}
\usepackage{amsfonts,amsmath,amssymb}
\usepackage{mathrsfs}
\usepackage{graphicx}
\usepackage{multirow}
\usepackage{epsfig}
\usepackage{graphicx}
\usepackage{subfigure}
\usepackage{booktabs}
\usepackage{array}
\usepackage{tabularx}
\usepackage{slashed}
\usepackage{braket}

\begin{document}
\baselineskip 20pt
\title{Hunting for sterile neutrino with future collider signatures}
\author{\vspace{1cm} Hao Yang$^1$\footnote[1]{yanghao2023@scu.edu.cn}, Bingwei Long$^{1,2}$\footnote[2]{bingwei@scu.edu.cn, corresponding author} and Cong-Feng Qiao$^{3,4}$\footnote[3]{qiaocf@ucas.ac.cn, corresponding author}\\}

\affiliation{
$^1$ College of Physics, Sichuan University, Chengdu, Sichuan 610065, China\\
$^2$ Southern Center for Nuclear-Science Theory (SCNT), Institute of Modern Physics, Chinese Academy of Sciences, Huizhou 516000, Guangdong, China\\ 
$^3$ School of Physics, University of Chinese Academy of Sciences, Yuquan Road 19A, Beijing 100049\\
$^4$ CAS Key Laboratory of Vacuum Physics, Beijing 100049, China
}

\begin{abstract}
We study the feasibility to observe sterile neutrino at the high energy colliders with direct production channels through $e^+e^-$, $ep$ collision, and indirect production channels through decays of heavy meson, baryon and Higgs. For $e^+e^-$ collision, the $e^+e^-\to\bar{\nu}_e N$ channel is explored with new signal selection method which tends to be efficient for light $m_N$, the constraints of active-sterile mixing $|U_{eN}|^2$ at the SuperKEKB, CEPC and ILC are expected to reach better lower limits than current experiments. For $ep$ collision, We investigate the heavy sterile neutrino production through a new channel via proton bremsstrahlung, i.e., $e^-\gamma \to NW^-$, hundreds of GeV heavy sterile neutrino can be probe and new limit on mixing is given. For heavy hadrons decay, the lepton-number-violating decays of $\Lambda_c,\ \Xi_{c},\ \Xi_{cc}$ and $\Lambda_b$ are explored via an intermediate on-shell Majorana neutrino in GeV scale. The branching fractions and the constraints for $|U_{\ell N}|^2$ are given, and hence may put new limits on this mass region. The ${\rm Higgs} \to W\mu\mu\pi$ channel is also considered to test massive neutrino within Higgs sector.
\end{abstract}

\maketitle

\section{INTRODUCTION}
The experimental observation of neutrino oscillation has conclusively shown small but non-zero neutrino mass. Since neutrinos are massless in the Standard Model (SM), the mass origin has become an important portal to physics beyond the standard model (BSM). There are generally three theoretical hypotheses: see-saw mechanism \cite{Minkowski:1977sc,Mohapatra:1979ia,Foot:1988aq}, radiative generation mass \cite{Babu:1989fg} and extra-dimensions \cite{Arkani-Hamed:1998wuz}. Originally, the smallness is resorted to the presence of extra field in see-saw models with far beyond electroweak energy mass scale. There are also models where extra fields are not so heavy \cite{Deppisch:2015qwa}, leaving the open possibility for sterile neutrinos in eV to TeV scale, and hence are feasible for collider searches. The reason for the smallness is yet not fully understood, but the very existence of neutrino mass may indicate the existence of a right-handed gauge-singlet (sterile) neutrino $N_R$. The Dirac or Majorana nature of $N_R$ can be identified by neutrinoless double-beta decay ($0\nu\beta\beta$) \cite{Doi:1985dx} of nucleus or other $W^{*,\pm}W^{*,\pm} \to \ell^{\pm}\ell^{\pm}$ induced processes, which led to lepton number violation (LNV) with $\Delta L=2$.

Various laboratory searches have put stringent constraints on sterile neutrino mixing with active ones in a broad mass range from eV to TeV. For the sterile neutrino below MeV, it is proposed to search kinks in the Kurie plots in the nuclear beta decays of $\rm ^{187}Re$ \cite{Galeazzi:2001py}, $\rm ^3H$ \cite{Hiddemann:1995ce}, $\rm ^{63}Ni$ \cite{Holzschuh:1999vy}, $\rm ^{35}S$ \cite{Holzschuh:2000nj}, $\rm ^{20}F$ \cite{Deutsch:1990ut}, $etc$. Analogously, search peaks in the energy spectra of charged pseudoscalar meson leptonic decays, $e.g.$, $\pi$ \cite{Britton:1992pg}, K \cite{Yamazaki:1984sj}. For heavier mass from MeV to GeV, sterile neutrino can be also tested via its effects on the lepton universality ratios BR($M^+\to \ell_1^+\nu_{\ell_1}$)/BR($M^+\to\ell_2^+\nu_{\ell_2}$) ($M=\pi,\ K,\ D,\ D_s$ \cite{Bryman:2019ssi, Bryman:2019bjg}) from their SM values, or via Majorana neutrino induced lepton number violation three/four body decay of heavy meson, $e.g.$, $D^+\to\pi^-(K^-)e^+e^+$ at the CLEO \cite{CLEO:2010ksb}, $B^+\to D^-+\ell^+\ell^{'+}$ at the Belle \cite{BELLE:2011bej}, $B^-\to\pi^+\mu^-\mu^-$ at the LHCb \cite{LHCb:2014osd} and $D\to K\pi e^+e^+$ at the BESIII \cite{BESIII:2019oef}. For the heavy sterile neutrino mass above GeV, using the possible production of sterile neutrino in the $Z^0$ boson decay $Z^0\to \nu(\bar{\nu}) N$, limits on the active-sterile mixing are obtained by L3 \cite{L3:1992xaz} and DELPHI \cite{DELPHI:1996qcc}, analogous production in the W boson decay $W\to\ell N$ is explored at CMS and ATLAS \cite{CMS:2022fut,ATLAS:2022atq}. For the mass above electroweak energy mass scale, direct searches were performed via same-sign dileptons plus jets \cite{CMS:2012wqj,CMS:2015qur,ATLAS:2015gtp,CMS:2018jxx}, $N\to\ell W$ \cite{L3:2001zfe,CMS:2018iaf}, or $N\to\ell jj$ \cite{ATLAS:2012ak}.

Now in this article, we will give several complementary investigations to the previous works by studying the feasibility of collider test for sterile neutrino through: (1) direct production via $e^+e^-,\ ep$ collision; (2) indirect production \footnote{Here one should not be confused with indirect effects to observables caused by off-shell sterile neutrino, e.g., the electroweak precision observables \cite{Blennow:2023mqx}. } via heavy particles decay, $e.g.$, Higgs, heavy meson/baryon. Several new lower $|U_{\ell N}|^2$ limits based on future experiments are set. 

The rest of the paper is organized as follows.
In Sec. II, we present the direct searches for sterile neutrino at the $e^+e^-,\ ep$ colliders.
In Sec. III, we investigate the indirect channels for sterile neutrino via heavy particles decay.
The last section is reserved for summary and conclusions.

\section{Direct production}
\subsection{$e^+e^-$ collision}
In the presence of one or several sterile neutrinos, active neutrinos in the flavor base are a mixture of the light and heavy sterile neutrinos in mass eigenstates. The lagrangian of interaction terms between sterile neutrino and gauge boson, Higgs boson in mass eigenstates are
\begin{align}
	-\mathcal{L} &= \dfrac{g}{\sqrt{2}}{W}^+_{\mu}\left(\sum_{\ell=e}^{\tau}\sum_{m=1}^{3}U_{\ell m}^{*}\bar{\nu}_m\gamma_{\mu}P_L\ell+\sum_{\ell=e}^{\tau}\sum_{m'=4}^{3+n}V_{\ell m'}^{*}\bar{N}_{m'}^c\gamma_{\mu}P_L\ell\right) + h.c.\nonumber\\		
	&+ \dfrac{g}{2\cos\theta_W}{Z}_{\mu}\left(\sum_{\ell=e}^{\tau}\sum_{m=1}^{3}U_{\ell m}^{*}\bar{\nu}_m\gamma_{\mu}P_L\nu_\ell+\sum_{\ell=e}^{\tau}\sum_{m'=4}^{3+n}V_{\ell m'}^{*}\bar{N}_{m'}^c\gamma_{\mu}P_L\nu_\ell\right) + h.c.\nonumber\\		
	&+ \dfrac{g\hspace{0.05cm}m_N}{2m_W}H\sum_{\ell=e}^{\tau}V^*_{\ell N}\bar{N^c}P_L\nu_{\ell} + h.c.,
	\label{EQLagrangian}
\end{align}
where $g = \frac{e}{\sin \theta_W}$, $\theta_W$ is the weak Weinberg mixing angle with $\sin^2\theta_W = 0.231$, $U_{\ell m}$ and $V_{\ell m}$ are the Pontecorvo-Maki-Nakagawa-Sakata (PMNS) matrix elements \cite{Pontecorvo:1957cp,Maki:1962mu}, the charge conjugated state is defined as $\psi^c = \mathcal{C} \bar{\psi}^{T}$, the left hand projection operator $P_L = \dfrac{1-\gamma^5}{2}$.

\begin{figure}[ht]			
	\centering
	\subfigure{\includegraphics[scale=0.6]{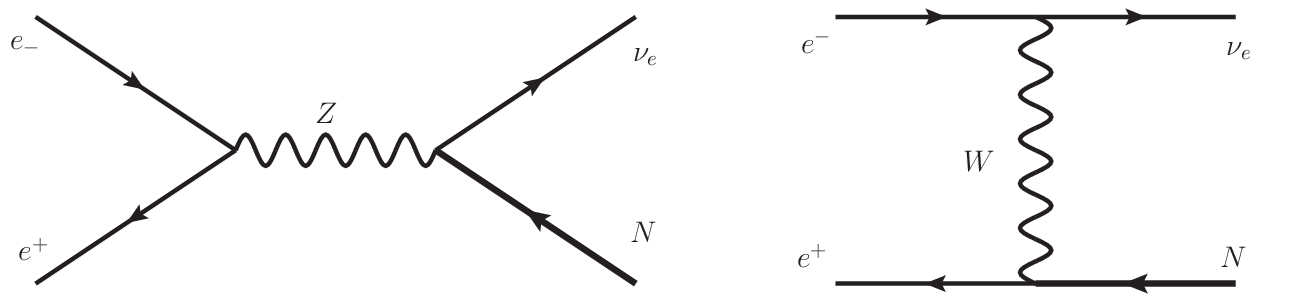}}
	\caption{Sterile neutrino production through Drell-Yan (Z) and W-exchange channels at the $e^+e^-$ collider.}
	\label{Figee2vN}
\end{figure}
In this section, we discuss the possible production of the sterile neutrino through $e^+e^-$ collision, and its subsequent decay. In this line, we address the dependency on active-sterile mixing $|U_{\ell N}|^2$ for the different colliders. There are generally two production channels, one is an annihilation channel through Z boson (s-channel), another is given by the exchange of a W boson (t-channel). We note that the sterile neutrino can be also produced via $e^+e^-$ annihilation into Higgs, which is largely suppressed by the tiny coupling between electron and Higgs.

Among the previous researches, various channels have been explored at next generation high energy lepton colliders, e.g., $e^+e^-$ \cite{Das:2012ze,Das:2018usr,Mekala:2022cmm,Das:2023tna}, $\mu^+\mu^-$ \cite{Mekala:2023diu}, $e^-e^-$ \cite{Banerjee:2015gca}, and see \cite{Antusch:2016ejd,Abdullahi:2022jlv} for review. In Refs. \cite{Das:2012ze,Banerjee:2015gca}, $\ell jj+\slashed{E}$ signal is proposed to reconstruct heavy neutrino N, and signals are required that the invariant mass of $\ell jj$ is near $m_N$ or the invariant mass of $jj$ is near $m_W$. In this work, we find that the cut on open angle ($\sum\theta_{\ell j_1j_2}=\theta_{\mu j_1}+\theta_{\mu j_2}+\theta_{j_1 j_2}$) would be another efficient way to exclude the background, especially for $m_N$ below $m_W$. For instance, the decay products $\mu j_1 j_2$ of a rest N (e.g., $m_N\sim\sqrt{s}$) tend to form a large open angle; while for light N (highly boosted), $\mu j_1j_2$ tend to stay along N flying direction, and hence form a small open angle. However, in the background channel, e.g., $e^+ e^- \to W^* W^* \to \mu \nu+j_1 j_2$, $\mu \nu$ and $j_1 j_2$ are produced by different W bosons, which is less energetic, hence the open angle between $\mu j_1j_2$ will be large. In this way, a open angle cut might be efficient way to select signals especially for light N. Aside from the investigation at the future high energy lepton colliders, the SuperKEKB can also provide unique test for sterile neutrino in this channel, which can put new lower limit for sterile neutrino mixing $|U_{eN}|^2$. 

Here, we study the sterile neutrino produced via s-channel and t-channel at the SuperKEKB, the Super Tau-Charm Facility (STCF), the Circular Electron Positron Collider (CEPC) and the International Linear Collider (ILC), where the final sterile neutrino is reconstructed by $\mu\pi$-channel for light N and by $\ell jj$-channel for heavy N. The analyses in this section and the rest sections are based on Feynman diagram through interaction vertex in (\ref{EQLagrangian}), we develop a private package to perform numerical integration and simulate kinematic of produced particle with the help of CUBA \cite{Hahn:2004fe}. Here and the rest of this article, one sterile neutrino N is supposed, one can easily extend our analysis to multi sterile neutrino models.

According to gauge-interaction lagrangian in (\ref{EQLagrangian}), the canonical matrix element square takes the form:  
\begin{align}	
	|\mathcal{M}(e^+e^-\to\bar{\nu}_eN)|^2/|U_{eN}|^2 &= 8G_F^2m_W^4\{\dfrac{4u(u-m_N^2)}{(m_W^2-t)^2}-\dfrac{4(2s_w^2-1)u(m_N^2-u)}{c_w^2(m_W^2-t)(m_Z^2-s)}\nonumber\\
	&+\dfrac{-m_N^2\left[4s_w^4(t+u)+u(1-4s_w^2)\right]+4s_w^4(t^2+u^2)+u^2(1-4s_w^2)}{c_w^4(m_Z^2-s)^2} \},
\end{align}
where $G_F$ is the weak interaction Fermi constant with $G_F = \frac{\alpha\pi}{\sqrt{2}m_W^2\sin^2\theta_W}$, the Mandelstam variables are define as $s = (p_{e^+}+p_{e^-})^2,\ t = (p_{e^-}-p_{\nu_e})^2,\ u = (p_{e^-}-p_N)^2$ with Mandelstam relation $s + t + u = m_N^2$, and $s_w=\sin\theta_W, c_w=\cos\theta_W, m_W=80.377\ {\rm GeV}, m_Z=91.187\ \rm GeV$. We have $|\mathcal{M}(e^+e^-\to\bar{\nu}_eN)|^2 = |\mathcal{M}(e^+e^-\to\nu_e\bar{N})|^2$ for the charge conjugated process.

The cross section is straightforward
\begin{equation}
	\sigma(e^+e^-\to\bar{\nu}_eN) = \dfrac{1}{2}\dfrac{1}{2}\dfrac{1}{2s}\dfrac{1}{8\pi s}\int_{m_N^2-s}^{0}|\mathcal{M}(e^+e^-\to\bar{\nu}_eN)|^2 dt,
\end{equation}
where the first two $\dfrac{1}{2}$ are spin-polarization average factors of electron and positron, $\dfrac{1}{2s}$ and $\dfrac{1}{8\pi s}$ are the flux and two-body final state phase space factor.

\begin{table}[ht]
	\caption{The center-mass energies and integrated luminosities of current and future $e^+e^-$ colliders. The integrated luminosity is estimated by $\rm 10^{34}\ cm^{-2}s^{-1} \sim 1\ ab^{-1}$ for 10 years operation.}
	\begin{center}
		\begin{tabular}{|p{2.5cm}<{\centering}|p{2.5cm}<{\centering}|p{2.5cm}<{\centering}|p{2.5cm}<{\centering}|p{2.5cm}<{\centering}|}
			\toprule
			\hline
			Collider                              & STCF & SuperKEKB & CEPC & ILC       \\
			\hline
			$\sqrt{s}$\ (GeV)                       & 7    & 10.6      & 250  & 500  \\
			\hline
			$\int d\mathcal{L}\ (\rm ab^{-1})$ & 5    & 80        & 3    & 1.8  \\ 
			\hline
		\end{tabular}
	\end{center}
	\label{TabsLuminosity}
\end{table}

The direct search for sterile neutrino is considered at the future lepton colliders, the SuperKEKB, the STCF, the CEPC and the ILC, each with its own physics focus on bottom, tau-charm, Higgs and Z respectively. The center-mass energies and integrated luminosities after 10 years operation are listed in TABLE \ref{TabsLuminosity}. For the same strength of mixing $U_{eN}$, t-channel is enhanced approximately 1$\sim$2 magnitudes compared with s-channel and hence get better sensitivity for $|U_{eN}|^2$. We note that the s-channel can be largely enhanced at the Z-pole running for the CEPC and the ILC, the mainly contribution can be regarded as on-shell Z boson decay \cite{DELPHI:1996qcc}, and $|U_{\mu N}|^2$ can be constrained via $e^+e^-\to Z^* \to N\nu_{\mu}$ \cite{Ding:2019tqq,Shen:2022ffi,Blondel:2022qqo}. 

As estimated in \cite{Zhang:2021wjj,Atre:2009rg}, the total decay width for Dirac sterile neutrino is set to be
\begin{equation}
	\Gamma_N^{Dirac} = \left\{ 
	\begin{aligned}
	    & 5 \sum_{\ell=e,\mu\tau}|U_{\ell N}|^2\dfrac{G_F^2m_N^5}{96\pi^3}   \hspace{0.3cm} m_N < m_W\\
	    & \sum_{\ell=e,\mu\tau}|U_{\ell N}|^2 \dfrac{3G_Fm_N^3}{16\pi\sqrt{2}} \hspace{0.33cm} m_N > m_W,
	\end{aligned}\right.
    \label{EqWidth}
\end{equation}
where  $U_{eN} \approx U_{\mu N} \approx U_{\tau N}$ is set for the universality consideration in the practice computation, see Refs.\cite{Helo:2010cw,Milanes:2016rzr,Zhang:2021wjj}. For Majorana sterile neutrino, $\Gamma_N^{Majorana} \approx 2\Gamma_N^{Dirac}$ is proposed.

\begin{figure}[htbp!]			
	\centering
	\subfigure{\includegraphics[scale=0.4]{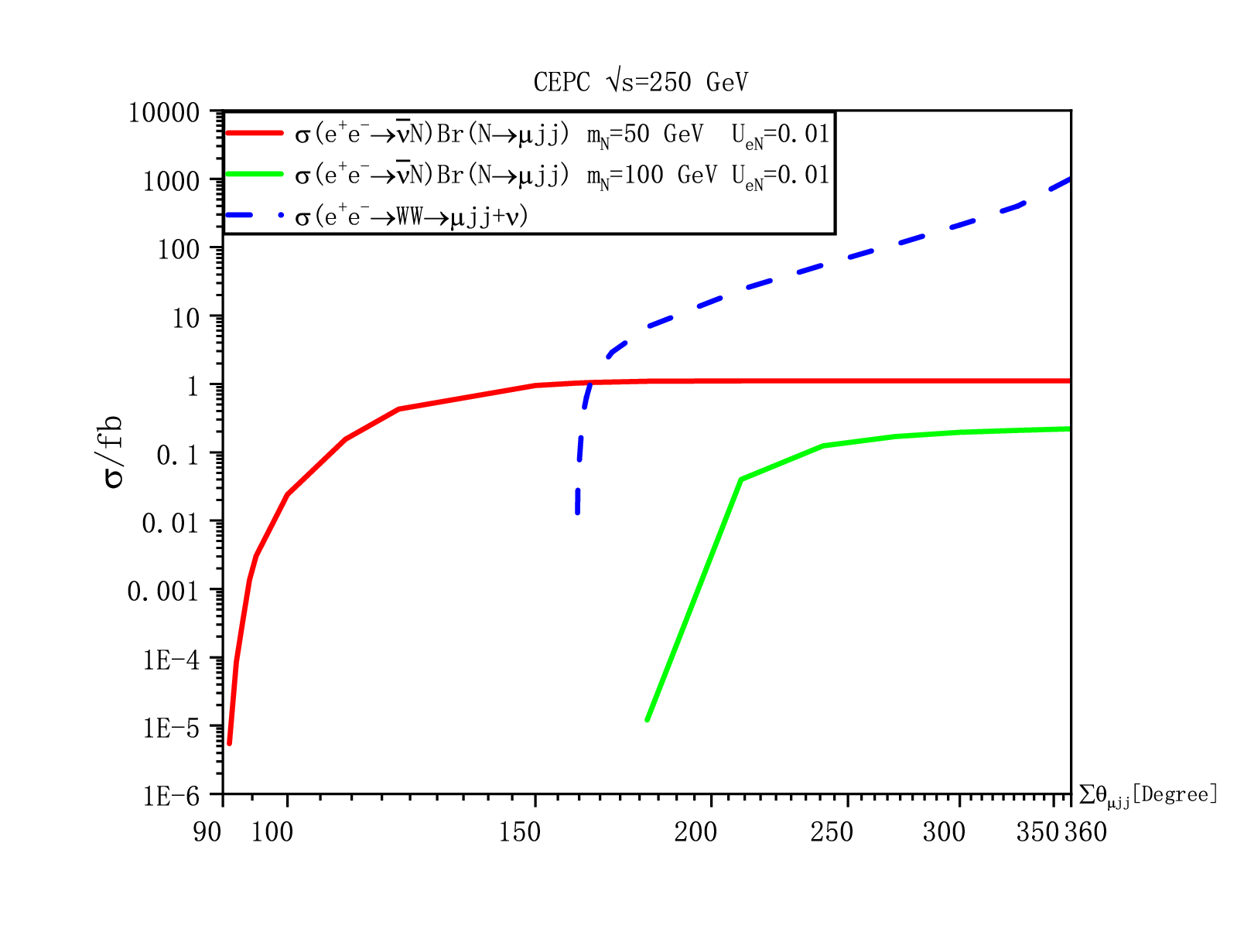}}
	\caption{The cross sections of sterile neutrino N production in Drell-Yan(Z) and W-exchange channels at the $e^+e^-$ collider and the background $e^+e^-\to WW \to \mu jj+\slashed{\nu}$ in different open angle cut $\sum \theta_{\mu jj}$. Here the basic cut is adopted and the charge conjugate is proposed.}
	\label{Figee2vNtheta}
\end{figure}

\begin{table}[htbp!]
	\caption{The cross sections for background $\sigma(e^+e^-\to WW \to\mu\bar{\nu}_\mu jj)$ at the CEPC with basic cut (BC) and two signal cuts (SC-I, SC-II) compared. The SC-I is energy variables cut: $|m_{\mu jj}-m_N| < 40\ \rm GeV$, which is adopted by Ref.\cite{Banerjee:2015gca}; the SC-II is open angle cut: $\sum_{\mu jj} < 160^\circ\ (270^\circ)$ for $m_N = 50\ (100)\ \rm GeV$ respectively, which is the signal selection strategy in this work.}
	\begin{center}
		\begin{tabular}{|p{4.5cm}<{\centering}|p{2.cm}<{\centering}|p{2.cm}<{\centering}|p{2.cm}<{\centering}|}
			\toprule
			\hline
			CEPC $\sqrt{s}=250\ \rm GeV$         & BC  &  BC+SC-I & BC+SC-II \\
			\hline
			$m_N = 50 \ \rm GeV $	& 1013 fb & 16.7 fb & 0 fb \\
			\hline 		
			$m_N = 100 \ \rm GeV $	& 1013 fb & 256 fb & 55 fb \\
			\hline			
		\end{tabular}
	\end{center}
	\label{Tabee2vNSCcompare}
\end{table}

\begin{table}[htbp!]
	\caption{The cross sections for $\sigma(e^+e^-\to \bar{\nu}N)Br(N\to\mu jj)$ and its background $\sigma(e^+e^-\to WW \to\mu\bar{\nu}_\mu jj)$ at the CEPC and the ILC after no cut, basic cut and signal cut. The background cross section without cut is in agreement with prediction in \cite{Aeppli:1993cb}. The $\theta_{min}$ is $160^\circ\ (270^\circ)$ for $m_N=50\ (100)\ \rm GeV$ at the CEPC; while for ILC, the $\theta_{min}$ is set to be $160^\circ\ (240^\circ)$ for $m_N=50\ (200)\ \rm GeV$ respectively.}
	\begin{center}
		\begin{tabular}{|p{4.0cm}<{\centering}|p{2.cm}<{\centering}|p{3.cm}<{\centering}|p{2.2cm}<{\centering}|p{4.2cm}<{\centering}|}
			\toprule
			\hline
			Cross Section         & No Cut  &  Basic Cut \ \ \ \ {\scriptsize $p_T^{\ell,j} > 10\ \rm GeV$, $|\eta^{\ell,j}| < 5, \slashed{E}_T > 10\ \rm GeV$} & Signal Cut\ \ \  {\scriptsize $\sum \theta_{\ell jj} < \theta_{min}$} & Settings \\
			\hline
			{\scriptsize $\sigma(e^+e^-\to \bar{\nu}N)Br(N\to\mu jj)$}   & 2.136 fb  & 1.101 fb & 1.02 fb & {\footnotesize \rm CEPC 250 GeV} \\
			{\scriptsize $\sigma(e^+e^-\to \mu jj + \slashed{\nu})$}     & 1.16 pb  & 1.01 pb & 0  & {\footnotesize $m_N=50\ \rm GeV\ U_{eN} = 0.01$}\\
			
			\hline
			{\scriptsize $\sigma(e^+e^-\to \bar{\nu}N)Br(N\to\mu jj)$}   & 0.29 fb   & 0.20 fb  & 0.12 fb & {\footnotesize \rm CEPC 250 GeV}\\
		    {\scriptsize $\sigma(e^+e^-\to \mu jj + \slashed{\nu})$}     & 1.16 pb  & 1.01 pb & 0.055 pb & {\footnotesize $m_N=100\ \rm GeV \ U_{eN} = 0.01$}\\
			\hline
			{\scriptsize $\sigma(e^+e^-\to \bar{\nu}N)Br(N\to\mu jj)$}   & 2.552 fb  & 1.808 fb & 1.807 fb & {\footnotesize \rm ILC 500 GeV}\\
			{\scriptsize $\sigma(e^+e^-\to \mu jj + \slashed{\nu})$}   & 0.50 pb  & 0.43 pb & 0  & {\footnotesize $m_N=50\ \rm GeV\ U_{eN} = 0.01$}\\
			\hline
			{\scriptsize $\sigma(e^+e^-\to \bar{\nu}N)Br(N\to\mu jj)$}   & 1.085 fb  & 1.012 fb & 0.645 fb & {\footnotesize \rm ILC 500 GeV}\\			
			{\scriptsize $\sigma(e^+e^-\to \mu jj + \slashed{\nu})$}   & 0.50 pb  & 0.43 pb & 0.016 pb & {\footnotesize $m_N=200\ \rm GeV \ U_{eN} = 0.01$}\\
			\hline
		\end{tabular}
	\end{center}
	\label{Tabee2vNcut}
\end{table}

\begin{figure}[htbp!]			
	\centering
	\subfigure{\includegraphics[scale=0.28]{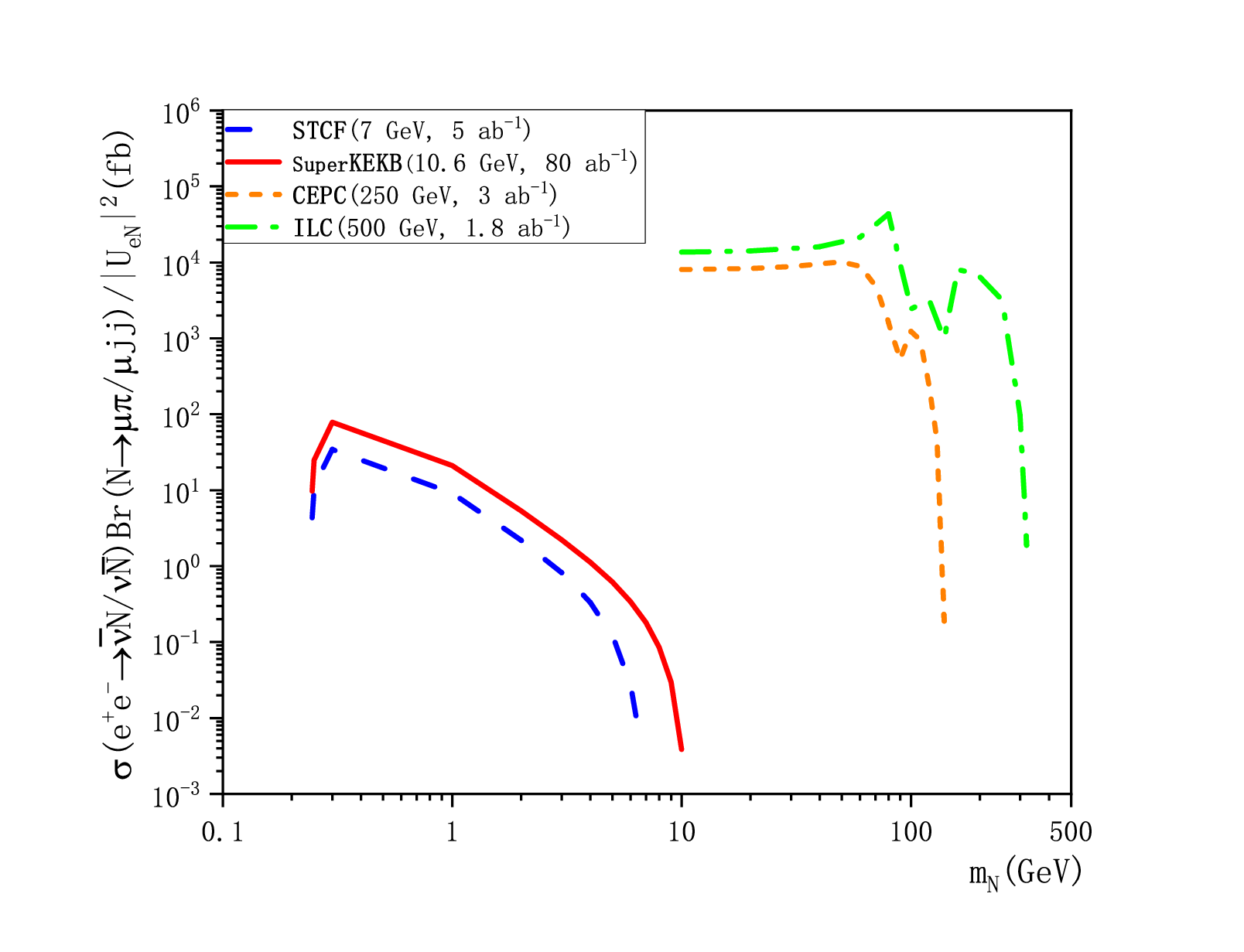}}
	\subfigure{\includegraphics[scale=0.28]{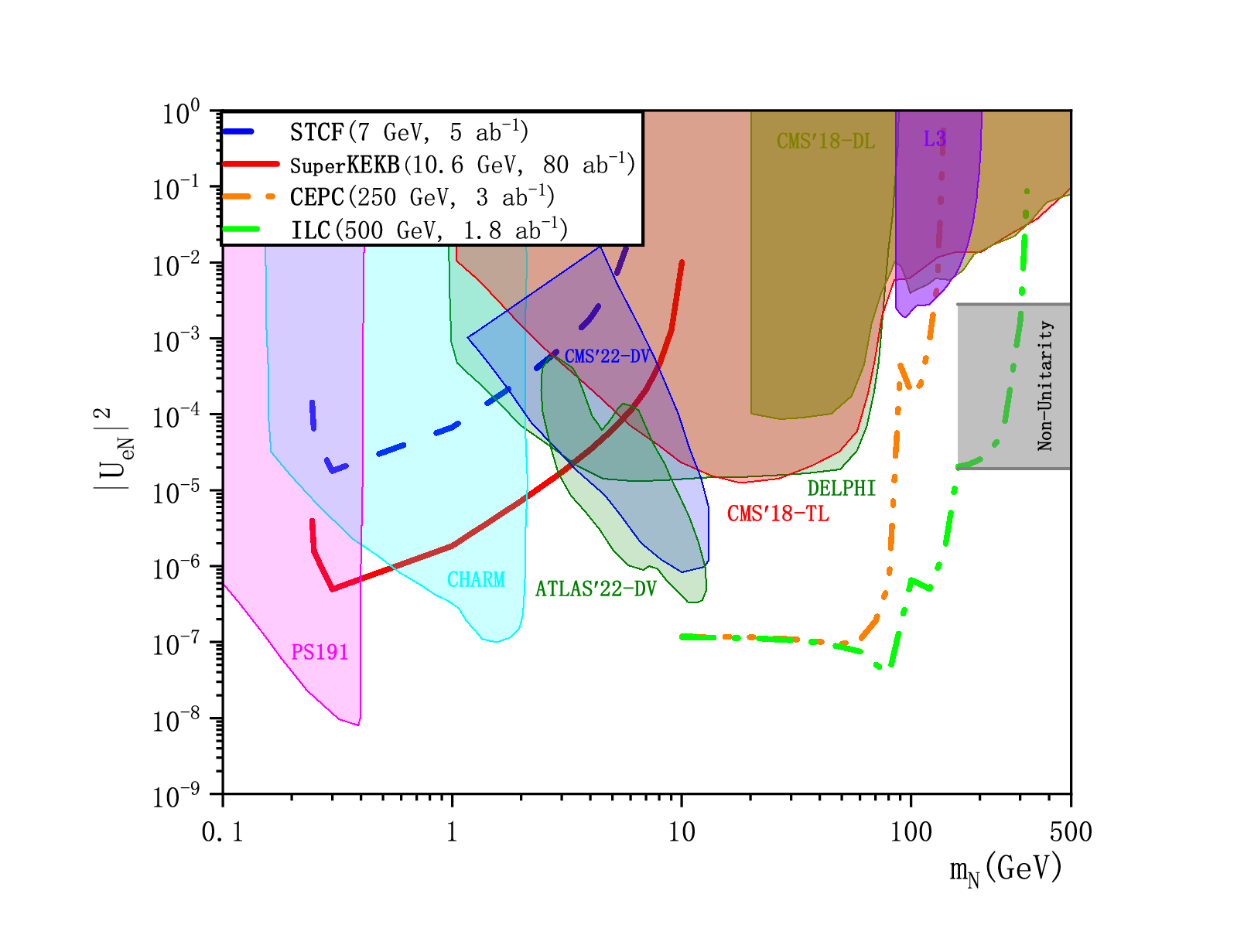}}
	\caption{The cross section of sterile neutrino N production in Drell-Yan(Z) and W-exchange channels at the $e^+e^-$ collider and the $|U_{eN}|^2$ sensitivity at the 95\% confidence level. Here for light mass, $N\to\mu\pi$ is used; while for heavy mass, $N\to\mu jj$ is adopted. The inflections are caused by signal cuts and definition of $\Gamma_N$. The current experiment limits labeled ``PS191'' \cite{Bernardi:1987ek} and ``CHARM'' \cite{CHARM:1985nku} are the constraints from beam-dump experiments; the constraint labeled ``DELPHI'' \cite{DELPHI:1996qcc} is given by $Z^0$ decay, the labels ``CMS$'$18-TL'' \cite{CMS:2018iaf} and ``CMS$'$18-DL'' \cite{CMS:2018jxx} are the trilepton and dilepton searches from direct N decay  at the CMS detector, the labels ``CMS$'$22-DV'' \cite{CMS:2022fut} and ``ATLAS$'$22-DV'' \cite{ATLAS:2022atq} are the displaced vertex searches at the CMS and ATLAS detector, and the label ``L3'' \cite{L3:2001zfe} is the result from $N_e\to e+W$ search at LEP. The label ``Non-Unitarity'' \cite{Blennow:2023mqx} is the constraints from non-unitarity effects in lepton mixing via dim-6 Weinberg operator at tree level.}
	\label{Figee2vNSigmaUeN}
\end{figure}

At the STCF and the SuperKEKB, we construct the signal via $N\to\mu\pi$ plus missing energy, the main background comes from $e^+e^-\to W^*W^*\to \mu\bar{\nu}_\mu\pi$ which is negligible due to double weak coupling at low center-of-mass energy \footnote{The cross section $\sigma(e^+e^-\to W^*W^*\to \mu\bar{\nu}_\mu\pi)$ is believe to be much small than $\sigma(e^+e^-\to W^*W^*\to \mu\bar{\nu}_\mu u\bar{d})$, which is 0.03(0.014) ab in SuperKEKB (STCF).}, hence we neglect this background in our analysis. At the CEPC and the ILC, the signal is composed of $\mu jj$ plus missing energy, and there tend to be small open angle between the signal lepton and jets for more energetic N, while the open angle is relative large for the background $e^+e^-\to WW\to \mu\bar{\nu}_\mu jj$ because lepton and jets are generated by different W bosons. In the previous researches, e.g., Ref.\cite{Banerjee:2015gca,Ding:2019tqq}, invariant mass of $\mu jj$ is applied for signal selection which will make a $m_N$ dependent background, one may analysis each proposed $m_N$ values to find trails of sterile neutrino. As we show in FIG.\ref{Figee2vNtheta}, the open angle cut may provide another strategy without reanalysis the background in each $m_N$ value. This strategy is efficient on low mass region (if $m_N$ is tens of GeV) compared with invariant mass selection, see TABLE \ref{Tabee2vNSCcompare} for detailed comparison. Hence we impose the open angle cut for signal selection at the CEPC (ILC): $\sum\theta_{\ell jj} < 160^\circ$ for $m_N < 80\ (150)\ \rm GeV$, the background can be excluded; $\sum\theta_{\ell jj} < 270^\circ\ (240^\circ)$ for $m_N > 80\ (150)\ \rm GeV$ with the background is reduced to 0.055(0.016) pb. The basic cuts are also imposed for the analysis at the CEPC and the ILC: $p_T^{\ell,j} > 10\ \rm GeV$, $|\eta^{\ell,j}| < 5$ and $\slashed{E}_T > 10\ \rm GeV$.  

According to our numerical results, see FIG.\ref{Figee2vNSigmaUeN}, the canonical cross sections are sub-fb to tens of pb with the increase of $\sqrt{s}$ from the STCF to the ILC. Considering that the integrated luminosities of those facilities are large, the signatures for GeV sterile neutrino can be well explored and hence make a constraint for active-sterile mixing $|U_{eN}|^2$. The constraint is estimated by solving $\frac{N_s}{\sqrt{N_s+N_B}} \approx 1.7$, which means signal significance at the 95\% confidence level \cite{Ding:2019tqq}, where $N_{s/B}$ is the signal/background events number. At the STCF, the center-of-mass energy may reach 7 GeV, the lower-limit of $|U_{eN}|^2$ will be $10^{-3}\sim10^{-4}$ at $0.3\sim2$ GeV with integrated luminosity of 5 $\rm ab^{-1}$. As for the SuperKEKB, the constraints can be further extended due to its high luminosity, and hence puts new complementary limits to current experiments. At the high energy electron-positron colliders, e.g., the CEPC and ILC, the physics potential for sterile neutrino can be extended to hundreds of GeV with $|U_{eN}|^2$ sensitivities of $10^{-3}\sim10^{-6}$, which will give better lower limits than ``CMS$'$18-TL'' \cite{CMS:2018iaf}, ``CMS$'$18-DL'' \cite{CMS:2018jxx}, ``L3'' \cite{L3:2001zfe} and ``Non-Unitarity'' \footnote{In the analysis of Ref.\cite{Blennow:2023mqx}, $m_N$ dependence has been omitted, hence the constraints might be applicable for $m_N \gg m_W$. The up(low)-bound is adopted from the maximal(minimal) constraints based on different neutrino schemes, i.e., 2N-SS (2 neutrino case), 3N-SS (3 neutrino case) and G-SS (general neutrino case). } constraint as indicated by previous research \cite{Das:2012ze,Das:2018usr,Mekala:2022cmm,Das:2023tna}.

\subsection{$ep$ collision}
Heavy sterile neutrino can be also probe at the future electron proton (Ion) colliders, like the Large Hadron electron Collider (LHeC) and the Electron-Ion Collider (EIC). There are extensive investigations on this topic through $e^-+q\to N+q'$ channel within W-exchange \cite{Lindner:2016lxq,Das:2018usr,Batell:2022ogj}. As indicated in Refs. \cite{Li:2018wut,Das:2023tna}, where the $\gamma$-$W^*$ mechanism is proposed using electron bremsstrahlung, the $\gamma$-$W^*$ channel through proton bremsstrahlung can also provide tests on heavy sterile neutrino for its relative clear signatures and negligible backgrounds. 

\begin{figure}[htbp!]			
	\centering
	\subfigure{\includegraphics[scale=0.3]{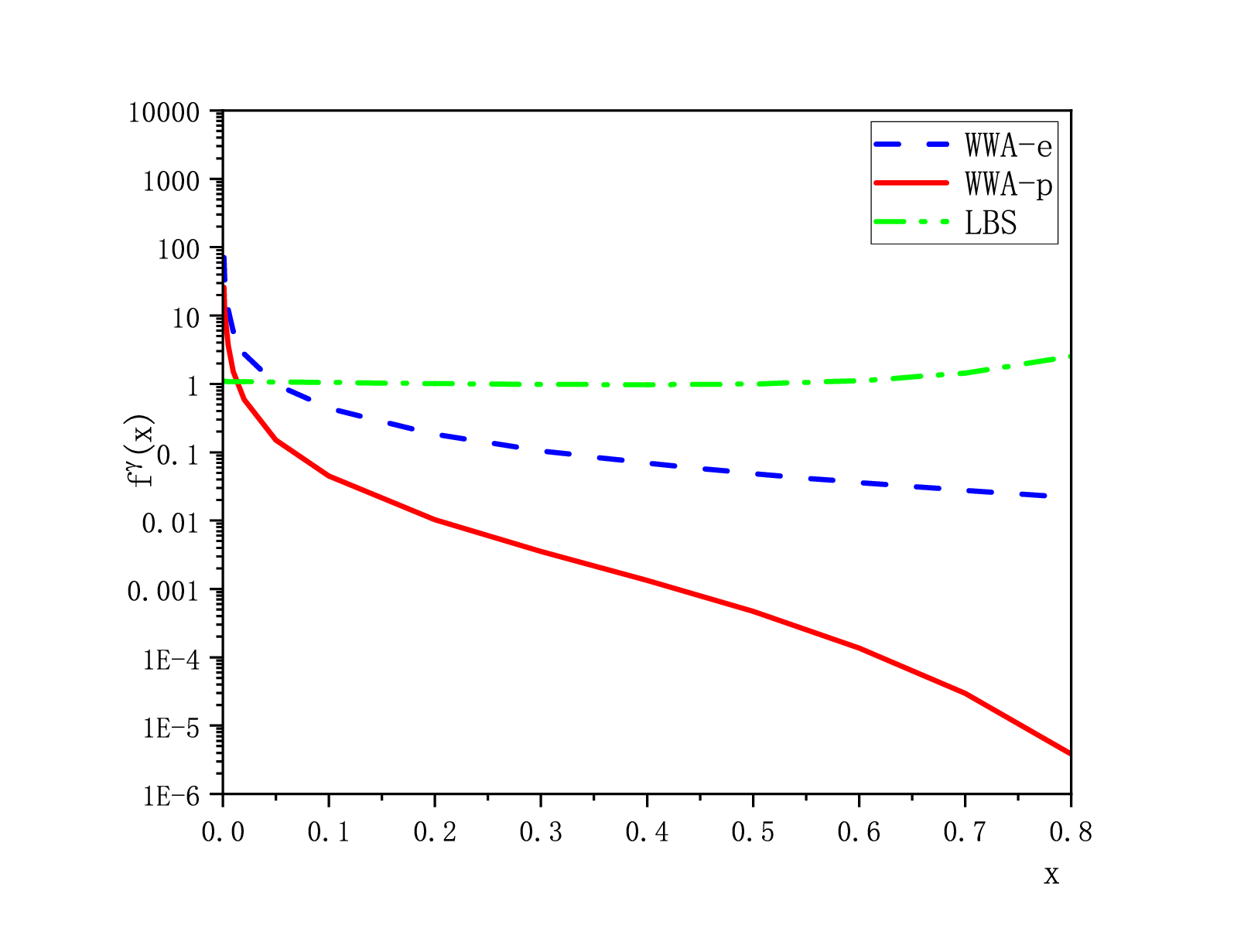}}
	\caption{The distribution functions of photon through electron/proton bremsstrahlung (WWA-e/p) and laser back scattering (LBS).}
	\label{FigphotonPDF}
\end{figure}

In this section, we explore the production mechanism of sterile neutrino in the context of  $\gamma$-$W^*$ interaction at the future $ep$ colliders, where the photon is produced via proton bremsstrahlung \cite{Gluck:2002fi}, the distribution function of photon is given in FIG.\ref{FigphotonPDF}. The photon from proton bremsstrahlung tends to less energetic compared with electron bremsstrahlung one, while the collider signature is relative clear due to the unbroken proton (several light jets are generated via broken proton and the signal jets could get submerged in complex jets background). 

The energy spectrum of proton bremsstrahlung photon can be well formulated in Weizsacker-Williams approximation (WWA-p) \cite{Gluck:2002fi}
\begin{equation}
	f^{p}_{\gamma}(x) = \dfrac{\alpha}{2\pi}\dfrac{2}{x}\left\{ \left[1-x+\dfrac{x^2}{4}(1+4a+\mu_p^2)\right]I + (\mu_p^2-1)\left[1-x+\dfrac{x^2}{4}\right]I' - \dfrac{1-x}{z^3}\right\},
\end{equation}
where $\mu_p = 2.79,\ a = 4.96$, $z=1+\dfrac{a}{4}\dfrac{x^2}{1-x}$ and
\begin{align}
	I  &= -\ln(1-\dfrac{1}{z}) - \dfrac{1}{z} - \dfrac{1}{2z^2}- \dfrac{1}{3z^3} \nonumber\\
	I' &= -\dfrac{1}{(a-1)^4}\ln(1+\dfrac{a-1}{z}) + \dfrac{1}{(a-1)^3z} - \dfrac{1}{2(a-1)^2z^2} + \dfrac{1}{3(a-1)z^3}. \nonumber
\end{align}

The total cross section can be expressed as the convolution of the $e^-+\gamma \to N + W^-$ cross section with the photon distribution function,
\begin{align}
	&\sigma(ep\to NW^{-}p) = \int dx f^{p}_{\gamma}(x) \hat{\sigma}(e^-\gamma \to NW^-)\\
	&\hat{\sigma}(e^-\gamma \to NW^-) = \dfrac{1}{2}\dfrac{1}{2}\dfrac{1}{2s}\dfrac{1}{8\pi s}\int_{t^{-}}^{t^{+}}|\mathcal{M}(e^-\gamma \to NW^-)|^2 dt,
\end{align}
where the Mandelstam variables are define as $s = (p_{e^-}-p_{\gamma})^2, t = (p_{e^-}-p_N)^2$, $t^{\pm} = \frac{m_N^2+m_W^2-s\pm\sqrt{\lambda(s,m_N^2,m_W^2)}}{2}$ with the $\rm K\ddot{a}llen$ function $\lambda(x,y,z) \equiv (x-y-z)^2-4yz$; the first two $\dfrac{1}{2}$ are spin-polarization average factors of electron and photon; $\dfrac{1}{2s}$ and $\dfrac{1}{8\pi s}$ are the flux and two-body final state phase space factor.
\begin{figure}[htbp]			
	\centering
	\subfigure{\includegraphics[scale=0.4]{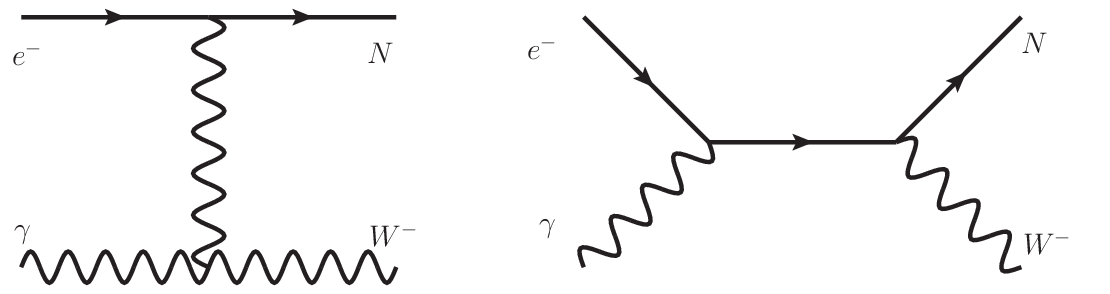}}
	\caption{The Feynman diagrams for heavy sterile neutrino production through $e^-\gamma \to NW^-$ channel, where the photon is produced via proton bremsstrahlung.}
	\label{Figer2NWFeyn}
\end{figure}

The amplitude square for $e^-\gamma \to NW^-$ is
\begin{align}
	|\mathcal{M}|^2 &= -\dfrac{|U_{eN}|^2 e^4}{2m_W^2s(m_W^2-t)^2s_w^2}\{ 2m_N^6(m_W^2-t) + m_N^4[6m_W^4-8m_W^2t+t(s+2t)]\nonumber\\            
	&+m_N^2[-6m_W^6-2m_W^4(3s+2t)+m_W^2(9s^2+14st+12t^2)+t(s^2-st-2t^2)]\nonumber\\
	&+4m_W^2[m_W^6-m_W^4(3s+t)+m_W^2(2s+t)^2-2s^3-4s^2t-3st^2-t^3]\}.
\end{align}

Here, we consider the production of sterile neutrino through $e^-\gamma \to NW^-$ channel at the future electron-proton collider, like the LHeC. For the events reconstruction, $N \to \mu jj$ channel is adopted, W boson is reconstructed via hadronic channel with $Br(W^-\to jj)$ $\approx 65\%$.
The LHeC is designed to reach a luminosity of $1.05\times10^{34}$, yields an approximately $1\ ab^{-1}$ integrated luminosity for 10 years operation. Furthermore, the Dirac or Majorana nature of the sterile neutrino can be identified via the sign of single lepton, $e.g.$, $N\to \ell^+ jj$ for only Majorana neutrino, while $N\to \ell^- jj$ for both Dirac and Majorana neutrinos.
\begin{figure}[ht]			
	\centering
	\subfigure{\includegraphics[scale=0.2]{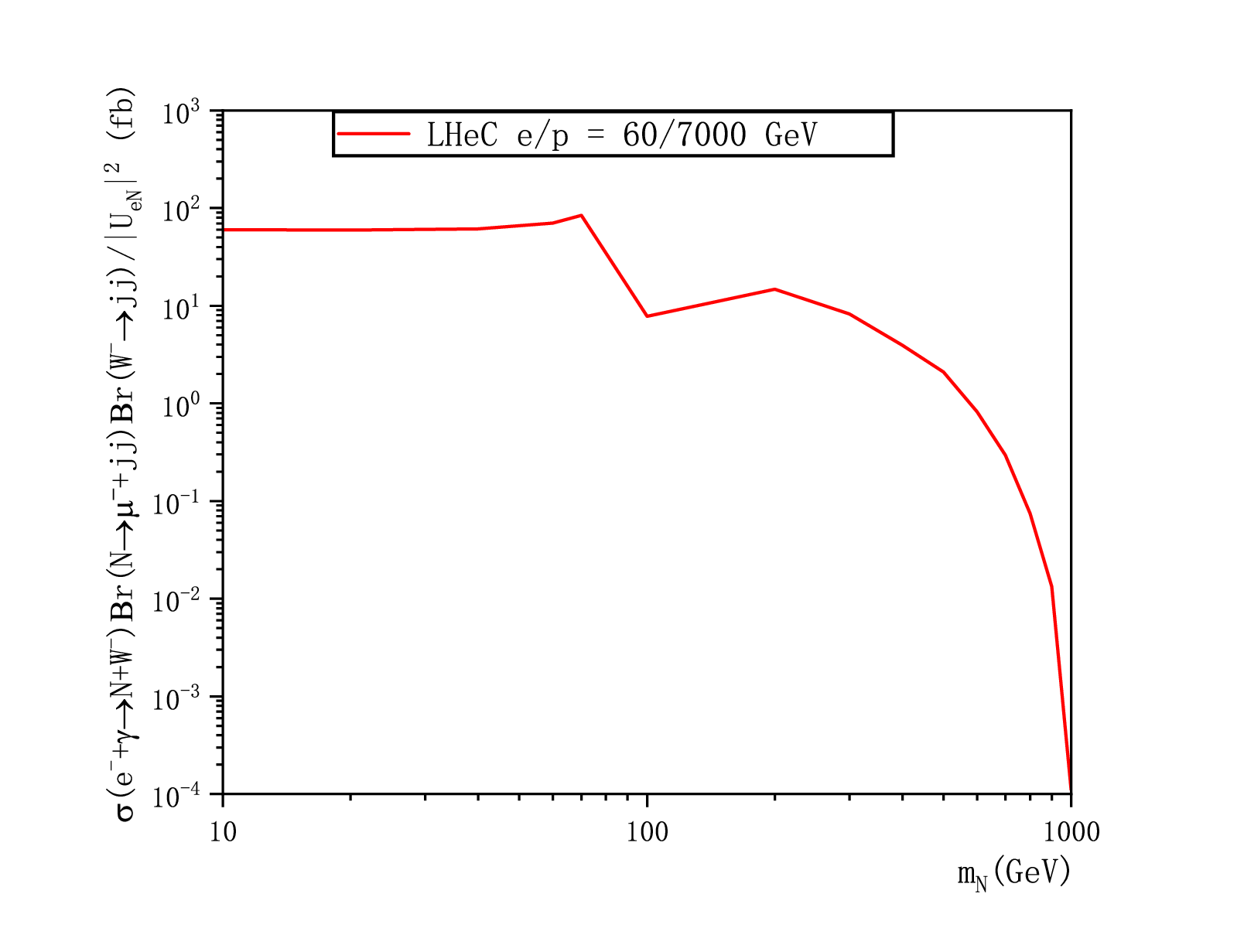}}
	\subfigure{\includegraphics[scale=0.2]{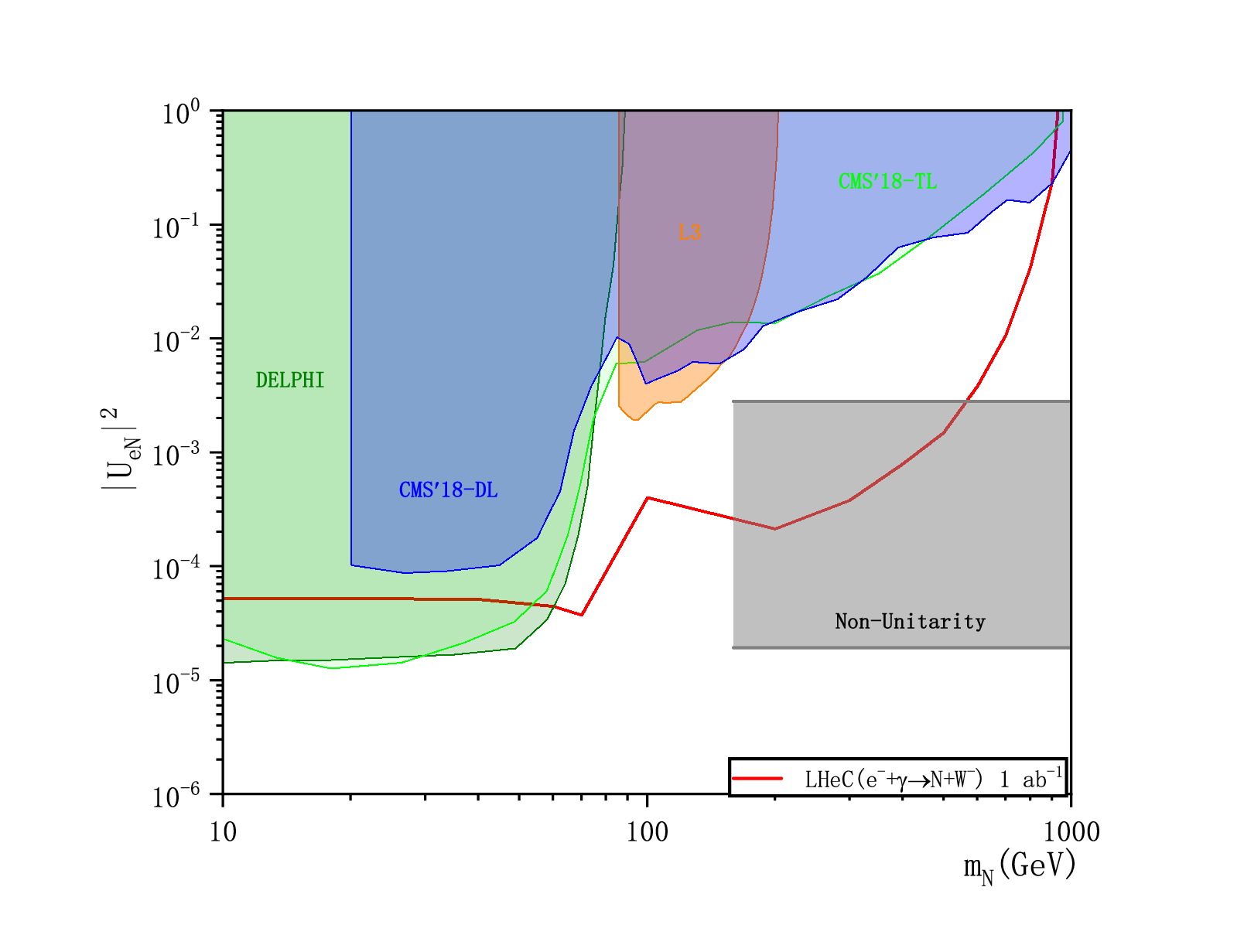}}	
	\caption{\textit{Left}: The canonical production cross section of $e^-\gamma \to NW^-$ at the LHeC with basic cuts. Here the photon is produced through proton bremsstrahlung. \textit{Right}: The sensitivities of sterile and active neutrino mixing $|U_{eN}|^2$ at the 95\% confidence level with integrated luminosity of $1\ \rm ab^{-1}$ at the LHeC. Here $N\to \mu^- jj$ and $W^-\to jj$ are taken into account for N and W reconstruction. The inflection near $m_W$ is caused by the definition of sterile neutrino total width. The current experiment constraint labeled ``DELPHI'' \cite{DELPHI:1996qcc} is given by $Z^0$ decay, the labels ``CMS$'$18-TL'' \cite{CMS:2018iaf} and ``CMS$'$18-DL'' \cite{CMS:2018jxx} are the trilepton and dilepton searches from direct N decay at the CMS detector, and the label ``L3'' \cite{L3:2001zfe} is the result from $N_e\to e+W$ search at LEP. The label ``Non-Unitarity'' \cite{Blennow:2023mqx} is the constraints from non-unitarity effects in lepton mixing via dim-6 Weinberg operator at tree level.}
	\label{Figer2NW}
\end{figure}

The incoming electron/proton beam energy at the LHeC is 60/7000 GeV, with center of mass energy $\sqrt{s} = 1296\ \rm GeV$. The photon is produced through high energy proton bremsstrahlung, therefore it provides a better condition for new physics phenomena searches due to its relative clear signatures  and small backgrounds. The main backgrounds are caused by $e^-+\gamma \to \nu_e + Z + W^-$ where Z decays into $\ell\bar{\nu}_\ell+jj$, and $e^-+\gamma \to \nu_e + W^- + W^- + W^+$ with W decays into leptons or jets. If the initial electron is lost in the beamline, as indicated by Ref.\cite{Blaksley:2011ey,Antusch:2019eiz}, it may also contribute to the background through process like $\gamma + \gamma\to Z + W^+ + W^-$, the two photons are produced via electron/proton bremsstrahlung. The cross sections for those backgrounds are below experiments tag limit 1/$\int\mathcal{L} \approx 1\ \rm ab$. In the next analysis, we adopt only the basic cuts (BC) for lepton and jets: $p_T^{\ell,j} > 10\ \rm GeV, |\eta^{\ell,j}| < 5$.

In FIG.\ref{Figer2NW}, the canonical cross section for $e^-\gamma \to NW^-$ and $|U_{eN}|^2$ sensitivity at the LHeC are given. According to our analysis, the canonical production cross section for $e^-\gamma \to NW^-$ will reach tens of fb if we probe a sterile neutrino in the electroweak energy mass scale. The cross section decreases to several fb for $m_N \sim 500\ \rm GeV$, and hence provides tests for heavy sterile neutrino in this region. The sensitivity of active-sterile neutrino mixing $|U_{eN}|^2$ with $100 < m_N < 500\ \rm GeV$ is estimated to be $10^{-3} \sim 10^{-4}$ level, therefore extends the current limit to a new level.

\section{Indirect Production}
\subsection{Kink search via B meson semileptonic decay}
\begin{figure}[ht]			
	\centering
	\subfigure{\includegraphics[scale=1.]{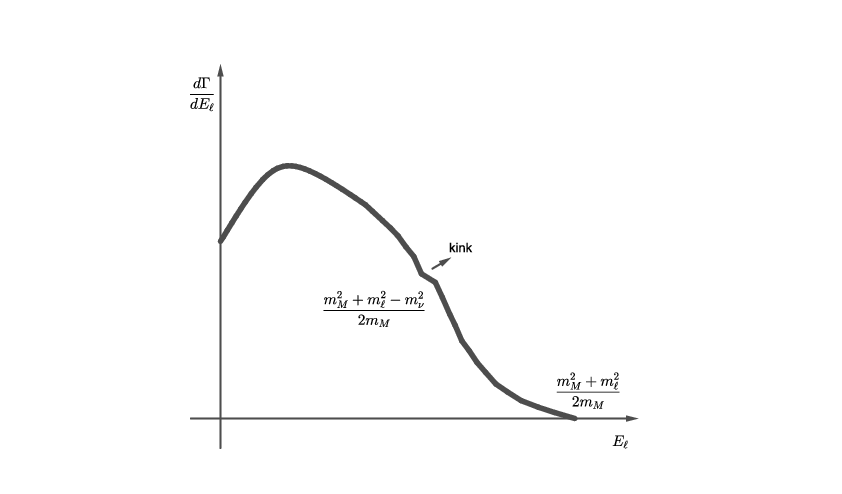}}
	\caption{Kink structure in lepton energy spectrum of $\beta$-decay for schematically review.}
	\label{FigKink}
\end{figure}

A massive sterile neutrino may be also investigated by searching kinks in lepton energy spectrum of the B meson semileptonic decay, see earlier nucleus $\beta$-decay spectrum kink search \cite{Shrock:1980vy}. The mechanism is straightforward, in three-body decay $B\to D+\ell\nu$, the maximal lepton energy at B meson rest frame is $E_\ell = \dfrac{m_B^2+m_\ell^2-(m_D+m_{\nu})^2}{2m_B}$. If the mass of sterile neutrino is in GeV, kink structure might exhibit at given energy point in the spectrum, see FIG.\ref{FigKink} for diagrammatic sketch. In the previous researches, focus are put on Nucleus or $\pi,K$ beam-dump experiments, in which $\pi,K$ are stopped, $e.g.$, by plastic scintillator inside a homogeneous magnetic field \cite{Daum:1977ec}. While at the SuperKEKB, B meson can be produced nearly at rest, thus may be applied to kink search.

The SuperKEKB, with designed luminosity of $8\times10^{35}\ {\rm cm^{-2}s^{-1}}$, will produce $2.7\times10^{10}\ \Upsilon(4S)$ [$\sigma(e^+e^-\to\Upsilon(4S)) = 1.08\ {\rm nb}$] per-year at $\Upsilon(4S)$ resonance \cite{Belle-II:2018jsg}.
The $B\bar{B}$ (approximately at rest) samples can be estimated by half the $\Upsilon(4S)$ yields, says $1.35\times10^{10} \ B\bar{B}$-pair. Considering an estimation for B meson semileptonic decay
\begin{equation}
	Br(B\to D+\ell N) \sim |U_{\ell N}|^2Br(B\to D+\ell \nu) = 2.2\times10^{-2}|U_{\ell N}|^2,
\end{equation}
and then the produced D meson is reconstructed via $K+n\pi$ channel, 
\begin{equation*}
	Br(D^+ \to K^-+2\pi^+) = 9.38\%,\ \ Br(\bar{D^0} \to K^-+\pi^+) = 3.94\%.
\end{equation*}

Therefore, the low sensitivity limit for $|U_{\ell N}|^2$ may reach $10^{-7}$ for mass region $1\sim4\ {\rm GeV}$. What deserves mentioning here is that the decay length of N may be long than detector size hence the directly reconstructed signal of N will suppressed, the direct search of N may be limited, so the kink method will provide another way to search sterile neutrino. A similar strategy can be used in the BES-III via $\Psi(3770)\to D\bar{D}$ where D meson is produced approximately at rest.

\subsection{Meson Decay}
\begin{figure}[ht]			
	\centering
	\subfigure{\includegraphics[scale=0.5]{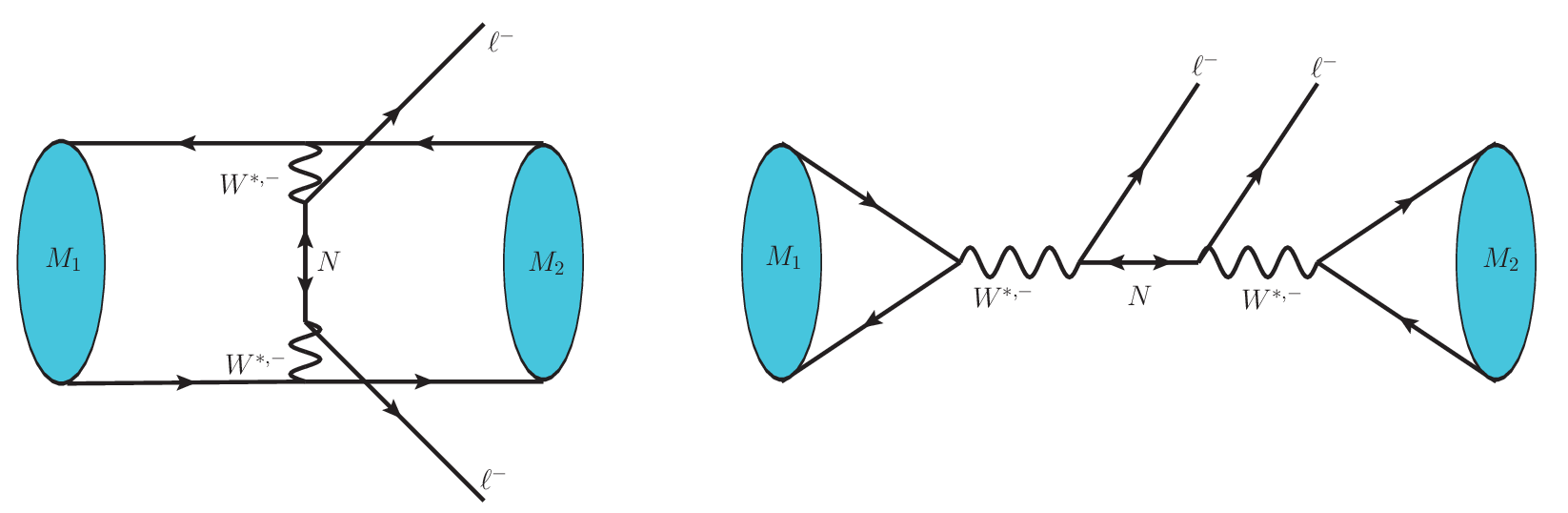}}
	\caption{$0\nu\beta\beta$ decay and sterile neutrino resonant decay of meson.}
	\label{FigM2Mll}
\end{figure}
The $0\nu\beta\beta$ decay ratios for charged meson ($\pi,K,D_{(s)},B_{(s,c)}$) are tiny, due to severe suppression either due to the small mass ratio like $\frac{m_N^2}{m_W^2}$ or small mixing $|U_{\ell N}U_{\ell' N}|^2$, hence are lack of experiment interest at present. For example, the $K^+ \to \pi^- + \mu^+\mu^+$ branching ratio for heavy sterile neutrino is \cite{Ali:2001gsa}:
\begin{align*}
	Br(K^+ \to \pi^{-} + \mu^+\mu^+) \sim 1.6\times10^{-12} (\dfrac{1\ GeV}{m_N})^2|U_{\mu N}|^4.
\end{align*}
Nevertheless, for a sterile neutrino lies between $m_{\pi}+m_{\mu} \le m_N \le m_K-m_{\mu}$, the branching ratio can be greatly enhanced by resonant effect \cite{Dib:2000wm,Shuve:2016muy,Mandal:2017tab,Chun:2019nwi}. In the narrow resonance approximation, the decay rate can be formulated as
\begin{equation}
	\Gamma(M_1 \to M_2 +\ell\ell) = \Gamma(M_1 \to \ell N)\dfrac{\Gamma(N \to \ell+M_2)}{\Gamma(N)}.
\end{equation}

In this way, the branching ratio can be easily estimated by the product of leptonic ratio of meson and $N\to \ell+M$ ratio, which is estimated to be several percents. The number of B/D meson at LHC is about $10^{12}/10^{13}$, thus provides experimental tests for sterile neutrino in mass region $1\sim4$ GeV.

\begin{table}[htbp!]
	\caption{The leptonic decay fraction of $K,D,D_s,B,B_c$, the $e^+\nu_e$ branching ratio are tiny due to small electron mass.}
	\begin{center}
		\begin{tabular}{|p{2.5cm}<{\centering}|p{2.5cm}<{\centering}|p{2.5cm}<{\centering}|p{2.5cm}<{\centering}|p{2.5cm}<{\centering}|p{2.5cm}<{\centering}|}
			\toprule
			\hline
			$Br$         & $K^+$\cite{ParticleDataGroup:2022pth}  &  $D^+$\cite{ParticleDataGroup:2022pth} & $D_s^+$\cite{ParticleDataGroup:2022pth} & $B^+$\cite{Belle:2019iji} & $B_c^+$\cite{Chang:1999gn}   \\
			\hline
			$\mu^+\nu_{\mu}$   & 0.635  & $3.74\times10^{-4}$ &$5.43\times10^{-3}$ &$4.3\times10^{-7}$ & $6.2\times 10^{-5}$ \\
			\hline
			$\tau^+\nu_{\tau}$ & -     & $1.2\times10^{-3}$   &$5.3\times10^{-2}$ 
			&$1.09\times10^{-4}$ & $1.47\times10^{-2}$\\ 
			\hline
		\end{tabular}
	\end{center}
	\label{TabBrM2munu}
\end{table}

\subsection{Baryon decay}
\begin{figure}[ht]			
	\centering
	\subfigure{\includegraphics[scale=0.5]{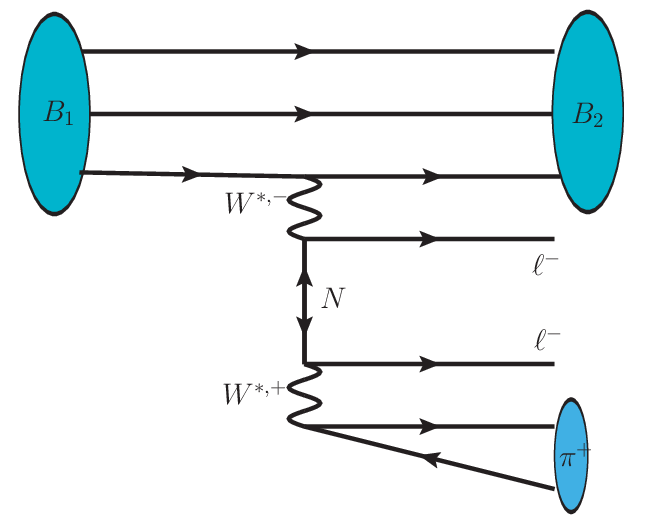}}
	\caption{Sterile neutrino resonant decay of heavy baryon.}
	\label{FigB2Bll}
\end{figure}
The resonant mechanism can be also applied in baryon semileptonic decay, $i.e.$, $B_1 \to B_2 +\ell N(\to\ell\pi)$, the  branching ratio is factorized as,
\begin{equation}
	Br(B_1 \to B_2 +\ell\ell+\pi) = \dfrac{\Gamma(B_1 \to B_2 +\ell N)}{\Gamma(B_1)}\dfrac{\Gamma(N \to \ell\pi)}{\Gamma(N)},
\end{equation}
where $\Gamma_{B_1(N)}$ is the total decay width of initial meson (N). The secondary decay width is well known,
\begin{equation}
	\Gamma(N \to \ell \pi) = \dfrac{G_F^2}{16\pi}|V_{ud}|^2|U_{\ell N}|^2f_{\pi}^2m_N\sqrt{\lambda(m_N^2,m_\ell^2,m_\pi^2)}[(1-\dfrac{m_\ell^2}{m_N^2})^2-\dfrac{m_\pi^2}{m_N^2}(1+\dfrac{m_\ell^2}{m_N^2})],
\end{equation}
where we set $m_{\pi}=139\ {\rm MeV}, f_{\pi}=130.4\ {\rm MeV}, m_{\mu}=105\ {\rm MeV}$ for numerical analysis. The Feynman diagram for this process is shown in FIG.\ref{FigB2Bll}, the amplitude can be formulated as
 \begin{equation}
 	\mathcal{M}(B_1 \to B_2+\ell N) = \dfrac{G_F}{\sqrt{2}}V_{q_1q_2}U_{\ell N} \bra{B_2}\bar{q}_2\gamma^\mu(1-\gamma^5)q_1\ket{B_1}[\bar{\ell}\gamma_\mu(1-\gamma^5)N],
 \end{equation}
where the CKM matrix elements are taken as: $V_{ud}=0.9737,\ V_{cs}=0.975,\ V_{cd}=0.221,\ V_{cb}=4.08\times10^{-2}$.

The hadronic transition matrix elements are parameterized in terms of six invariant form factors,
\begin{align}
	\bra{B_2}\bar{q}_2\gamma^\mu(1-\gamma^5)q_1\ket{B_1} &= \bar{u}(B_2)[\gamma^\mu f_1(q^2) + \sigma^{\mu\nu}q_{\nu}\dfrac{f_2(q^2)}{m_{B_1}} + q^{\mu}\dfrac{f_3(q^2)}{m_{B_1}}]u(B_1)\\ \nonumber
	&-\bar{u}(B_2)[\gamma^\mu g_1(q^2) + \sigma^{\mu\nu}q_{\nu}\dfrac{g_2(q^2)}{m_{B_1}} + q^{\mu}\dfrac{g_3(q^2)}{m_{B_1}}]\gamma^5u(B_1),
\end{align}
where $u(B_{1/2})$ are Dirac spinors of the initial/final baryon with $q = p_{B_1}-p_{B_2}$.

\begin{figure}[ht]			
	\centering
	\subfigure{\includegraphics[scale=0.5]{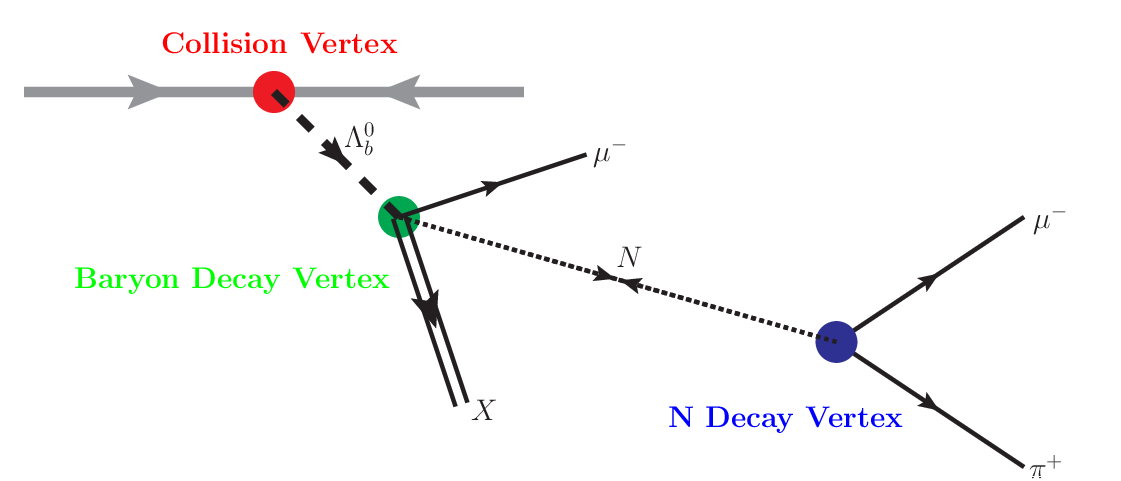}}
	\caption{The schematic event structure of sterile neutrino resonant semileptonic decay of $\Lambda_b^0$.}
	\label{FigB2Bllpi_V}
\end{figure}

Note that a considerable number of heavy baryons can be produced at the LHC, hence it may provide opportunity for experiment searches of the sterile neutrino. Experimental limits from the searches of $|\Delta L| = 2$ processes can be reinterpreted as $|U_{\mu N}|^2$ constraints versus $m_N$, with considering the detection efficiencies in TABLE \ref{TabChannelEfficiency}. Precise computation of the detection efficiencies requires fully simulated decay-specific Monte Carlo samples, that is reconstructed in the same manner as real data and with a simulation of the full detector, which is out of the range of this paper. We note that the decay width for N is proportional to $|U_{\ell N}|^2$ and $m_N^5$ in Eq.(\ref{EqWidth}), a light neutrino with small mixing will fly long before decay, hence the decaying probability of the neutrino within detector will be exponentially suppressed,
\begin{equation}
	P_N = 1-\exp[-\dfrac{L}{\tau_N \gamma_N\beta_N}],
\end{equation} 
where $\gamma_N=1/\sqrt{1-\beta_N^2},\ \beta_N = v_N/c$, $\tau_N$ is the lifetime of the neutrino, and L is the length of detector. Considering the rapidity and transverse momentum are 3 and 4 GeV within typical acceptance of LHCb detector, and the detector length is set to be 12 m \cite{LHCb:2014set} which requires N decays before the calorimeter system (It provides the identification of leptons and hadrons), the decay length effects are considered in the following analysis. In contrast, the sizable decay length between N production and decay vertexes will in turn improve the efficiency of events reconstruction \cite{CMS:2022fut}, if partially reconstruction strategy is adopted. For the fully reconstruction, as indicated in previous $\Lambda_{b/c}$ studies \cite{Mejia-Guisao:2017nzx,Zhang:2021wjj}, e.g., a complete $\Lambda_c$ event is required in $\Lambda_b\to \Lambda_c+\mu\mu\pi$ decay, while for partially reconstruction method, e.g., one need only tag a baryon (for instance proton produced in $\Lambda_c$ decay) despite its type. The two methods are compared in the following analyses. 
\begin{table}[ht]
	\begin{center}
		\caption{The masses, lifetimes, decay lengths \cite{ParticleDataGroup:2022pth} at the rest/lab frame of long live particles with $p_t = 4\ {\rm GeV},\ y = 3$.}	
		\centering
		\scalebox{0.7}{
			\begin{tabular}{|m{1.42cm}<{\centering}|m{2.1cm}<{\centering}|m{2.5cm}<{\centering}|m{2.5cm}<{\centering}|m{2.5cm}<{\centering}|}	
				\toprule
				\hline			
				particle       & mass (MeV) & lifetime ($\tau$, s)   & $c\tau$            & $\gamma\beta c\tau$ \\
				\hline
				$\mu$           & 105        & $2.19\times10^{-6}$    & $658.6\ \rm m$     & $2.52\times10^5\ \rm m$\\
				\hline
				$\tau$          & 1176       & $2.9\times10^{-13}$    & $87\ \mu \rm m$    & $2.15\ \rm mm$\\
				\hline
				$\pi^{\pm}$     & 139        & $2.6\times10^{-8}$     & $7.8\ \rm m$       & $2.26\times10^3\ \rm m$\\
				\hline
				$K^{\pm}$       & 493        & $1.23\times10^{-8}$    & $3.71\ \rm m$      & $3.05\times10^2\ \rm m$\\
				\hline
				$D^{\pm}$       & 1869       & $1.04\times10^{-12}$   & $311.8\ \mu \rm m$ & $7.41\ \rm mm$\\
				\hline
				$D^0$           & 1864       & $0.41\times10^{-12}$   & $122.9\ \mu \rm m$ & $2.93\ \rm mm$\\
				\hline
				$D_s^{\pm}$     & 1968       & $0.5\times10^{-12}$    & $149.9\ \mu \rm m$ & $3.41\ \rm mm$\\
				\hline
				$B^{\pm}$       & 5279       & $1.638\times10^{-12}$  & $491.1\ \mu \rm m$ & $6.18\ \rm mm$\\
				\hline
				$B^0$           & 5279       & $1.52\times10^{-12}$   & $455.7\ \mu \rm m$ & $5.74\ \rm mm$\\
				\hline
				$B_s^0$         & 5366       & $1.51\times10^{-12}$   & $452.7\ \mu \rm m$ & $4.54\ \rm mm$\\
				\hline
				$B_c^{\pm}$     & 6275       & $0.507\times10^{-12}$  & $152\ \mu \rm m$   & $1.81\ \rm mm$\\
				\hline
				$\Lambda$       & 1115      & $2.63\times10^{-10}$    & $7.89\ \rm cm$     & $2.96\ \rm m$\\
				\hline
				$\Sigma^+$      & 1189       & $0.8018\times10^{-10}$ & $2.404\ \rm cm$    & $0.85\ \rm m$\\
				\hline
				$\Sigma^-$      & 1197       & $1.479\times10^{-10}$  & $4.434\ \rm cm$    & $1.56\ \rm m$\\
				\hline
				$\Xi^0$         & 1314       & $2.9\times10^{-10}$    & $8.71\ \rm cm$     & $2.81\ \rm m$\\
				\hline
				$\Xi^-$         & 1321       & $1.639\times10^{-10}$  & $4.91\ \rm cm$     & $1.57\ \rm m$\\
				\hline
				$\Omega^-$      & 1672       & $0.821\times10^{-10}$  & $2.461\ \rm cm$    & $0.62\ \rm m$\\
				\hline
				$\Lambda_c^+$   & 2286       & $2.01\times10^{-13}$   & $59.9\ \mu \rm m$  & $1.21\ \rm mm$\\
				\hline
				$\Xi_c^+$       & 2467       & $4.42\times10^{-13}$   & $132\ \mu \rm m$   & $2.53\ \rm mm$\\
				\hline
				$\Xi_c^0$       & 2470       & $1.12\times10^{-13}$   & $33.6\ \mu \rm m$  & $0.64\ \rm mm$\\
				\hline
				$\Omega_c^0$    & 2695       & $6.9\times10^{-14}$    & $21\ \mu \rm m$    & $0.38\ \rm mm$\\
				\hline
				$\Xi_{cc}^{++}$ & 3621       & $2.56\times10^{-13}$   & $76.7\ \mu \rm m$  & $1.15\ \rm mm$\\
				\hline
				$\Lambda_b^0$   & 5619       & $1.464\times10^{-12}$  & $439.5\ \mu \rm m$ & $5.41\ \rm mm$\\
				\hline
				$\Xi_b$         & 5794       & $1.56\times10^{-12}$   & $467.7\ \mu \rm m$ & $5.70\ \rm mm$\\
				\hline
				$\Omega_b$      & 6046       & $1.57\times10^{-12}$   & $470.7\ \mu \rm m$ & $5.66\ \rm mm$\\						
				\hline
			\end{tabular}
		}
	\end{center}
	\label{TableMassLifetime}		
\end{table}

\begin{figure}[ht]			
	\centering
	\subfigure{\includegraphics[scale=0.35]{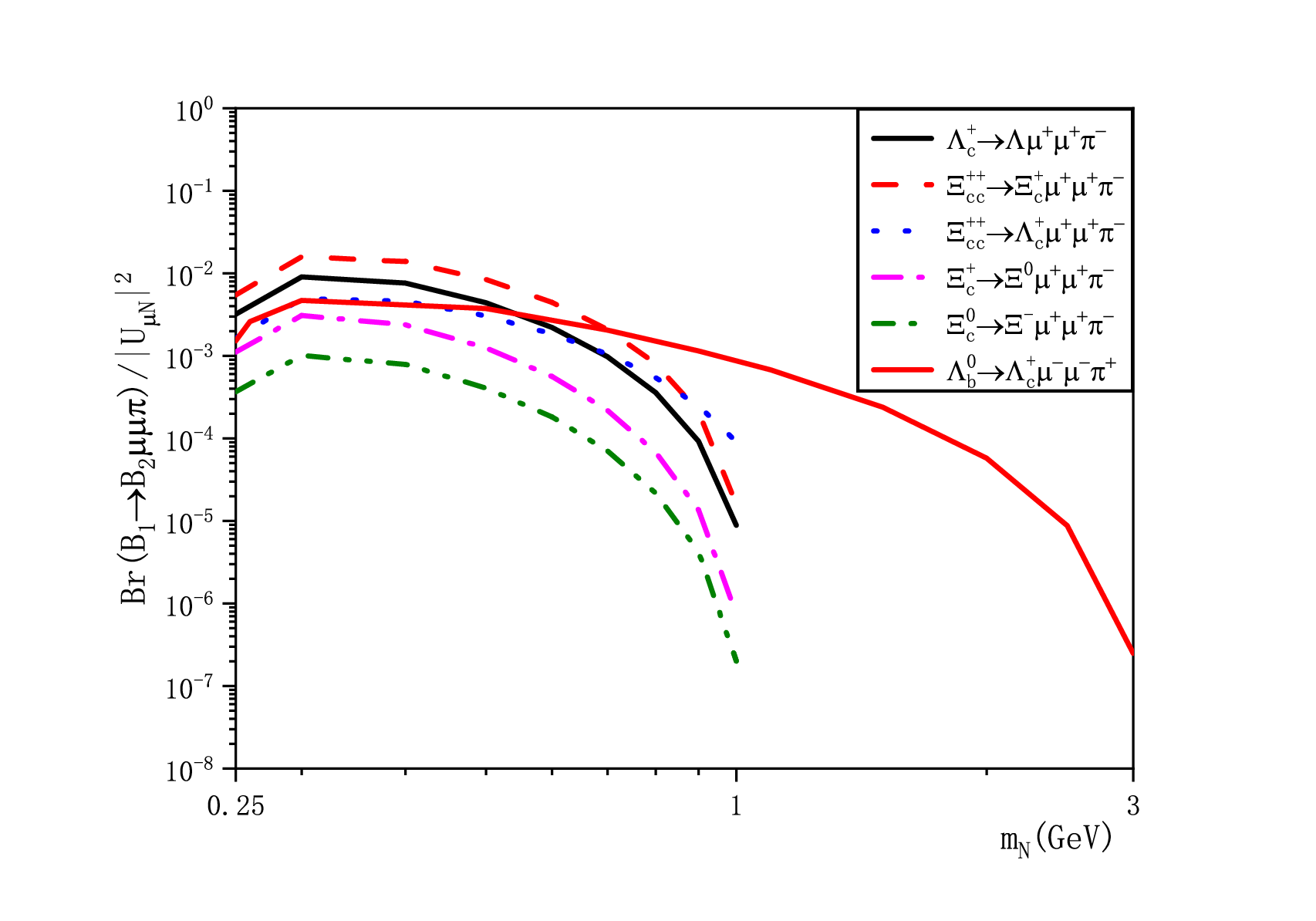}}
	\subfigure{\includegraphics[scale=0.27]{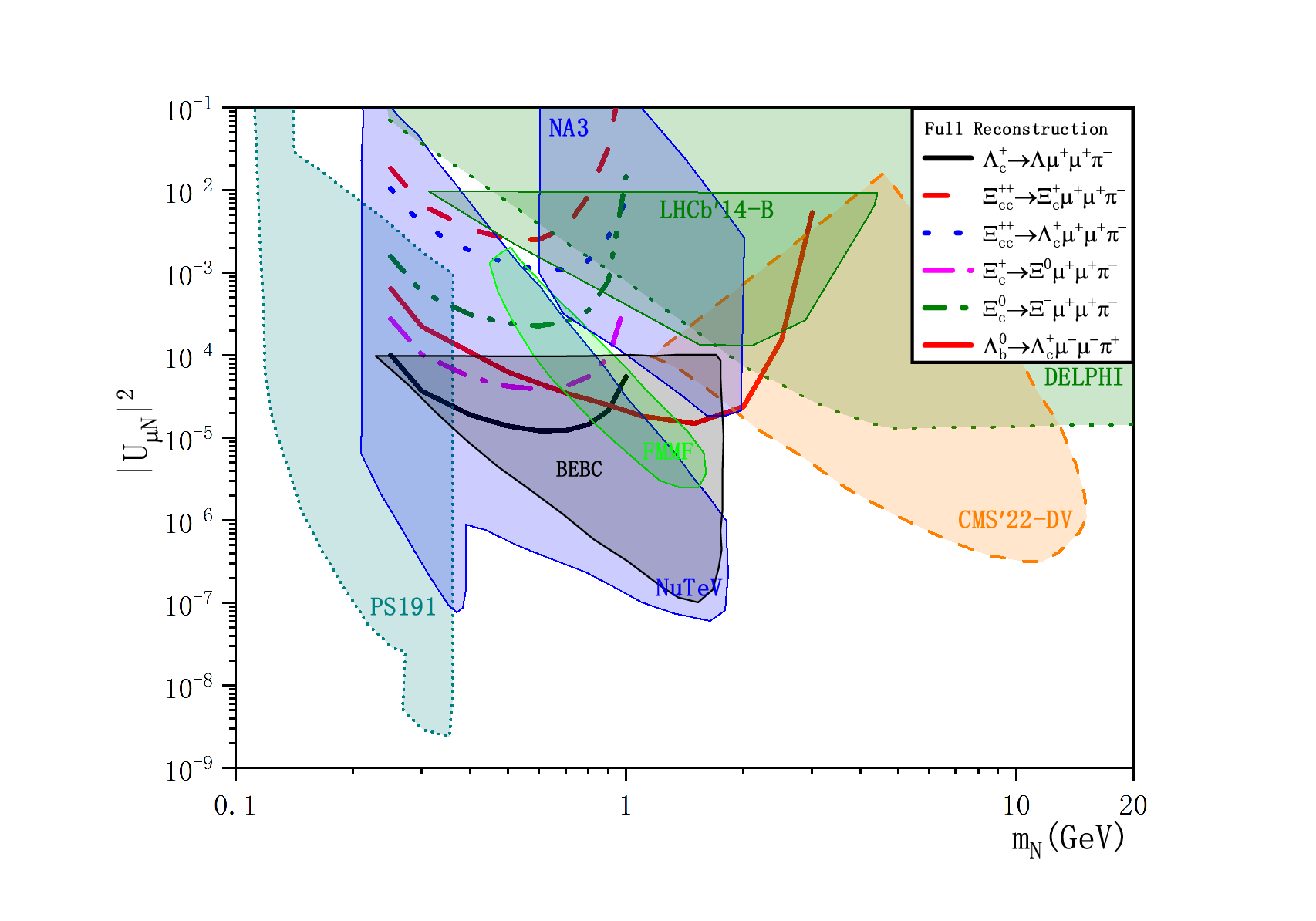}}
	\subfigure{\includegraphics[scale=0.27]{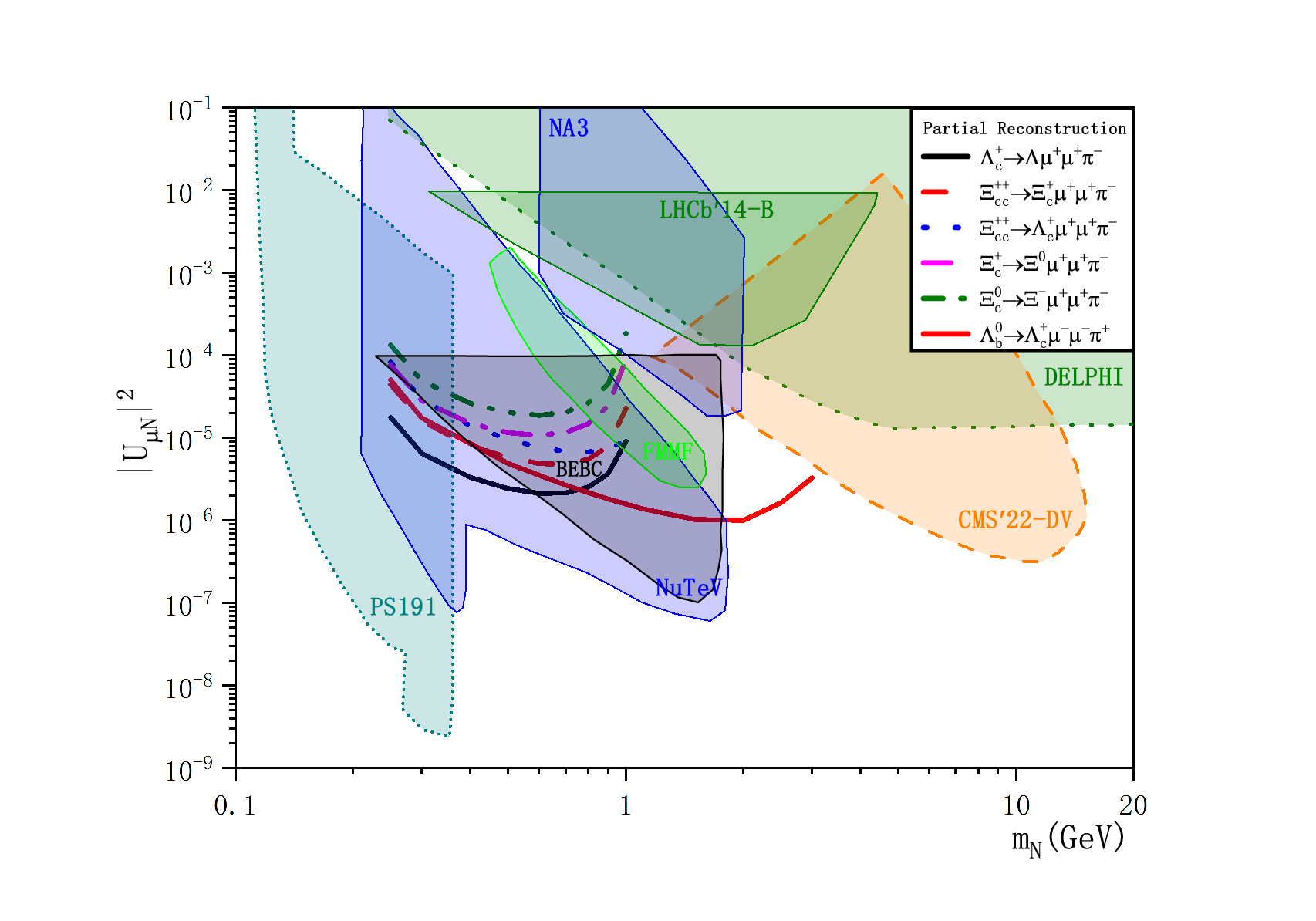}}
	\caption{The canonical branching fractions of sterile neutrino exchange four-body heavy baryon decay, and the sensitivities of $|U_{\mu N}|^2$ at the LHCb with an integrated luminosity of 50 $\rm fb^{-1}$ at the 95\% confidence level. The current experiment limits labeled ``PS191'' \cite{Bernardi:1987ek}, NA3 \cite{NA3:1986ahv}, BEBC \cite{WA66:1985mfx}, FMMF \cite{FMMF:1994yvb} are the constraints from beam-dump experiments, the label ``NuTeV'' \cite{NuTeV:1999kej} is the result from direct N decay search, the label ``LHCb$'$14-B'' is the constraints from B meson decay search, the constraint labeled ``DELPHI'' \cite{DELPHI:1996qcc} is given by $Z^0$ decay, and the labels ``CMS$'$22-DV'' \cite{CMS:2022fut} is the displaced vertex searches at the CMS detector.}
	\label{FigB2Bllpi}
\end{figure}

Apart from $\Lambda_{b/c}$, we also explore the four-body $|\Delta L| = 2$ decay of the $\Xi_{cc}^{++},\Xi_c^+,\Xi_c^0$ baryons,
\begin{align*}
	&\Xi_{cc}^{++} \to \Xi_c^+ + \mu^+\mu^+\pi^-,\\
	&\Xi_{cc}^{++} \to \Lambda_c^+ + \mu^+\mu^+\pi^-,\\
	&\Xi_{c}^{+} \to \Xi^0 + \mu^+\mu^+\pi^-,\\
	&\Xi_{c}^{0} \to \Xi^- + \mu^+\mu^+\pi^-,
\end{align*}  
via the exchange of Majorana neutrino with kinematically allowed mass, $m_{\mu}+m_{\pi} < m_N < m_{B_1}-m_{B_2}-m_{\mu}$. Within this mass region, the narrow width approximation \cite{Atre:2009rg} is valid due to $\Gamma_N \ll m_N$. For numerical evaluation, we use the form factors for $\Xi_{cc}^{++} \to \Xi_c^+$ \cite{Shi:2019hbf},  $\Xi_{cc}^{++} \to \Lambda_c$ \cite{Shi:2019fph}, $\Xi_c^+ \to \Xi^0$ \cite{Azizi:2011mw}, $\Xi_c^0 \to \Xi^-$ \cite{Azizi:2011mw} in the references respectively, the canonical branching fractions $\frac{Br(B_1 \to B_2 + \mu\mu\pi)}{|U_{\mu N}|^2}$ are given in FIG. \ref{FigB2Bllpi}.
\begin{table}[ht]
	\caption{The estimated production numbers of heavy baryons at the LHCb in 13 TeV with an accumulated luminosity of 50 $\rm fb^{-1}$.} 	
	\label{TabNBaryon}
	\begin{center}		
		\begin{tabular}{|p{2.5cm}<{\centering}|p{2.5cm}<{\centering}|p{2.5cm}<{\centering}|p{2.5cm}<{\centering}|p{2.5cm}<{\centering}|p{2.5cm}<{\centering}|}
			\toprule
			\hline
			& $\Lambda_c^+$  &  $\Lambda_b^0$ & $\Xi_{cc}^{++}$ & $\Xi_c^+$ & $\Xi_c^0$   \\
			\hline
			number($10^{12}$)    & 48.3          & 3.73          & 0.715           & 1.49     & 19.0 \\
			\hline
		\end{tabular}
	\end{center}    
\end{table}

At the LHC, the production number of heavy baryons can be estimated by fragmentation fractions of c-quark and b-quark with $f(c\to\Lambda_c^+) = 20.4\%$ \cite{ALICE:2021dhb}, $f(c\to\Xi_c^0) = 8\%$ \cite{ALICE:2021dhb}, $f(b\to\lambda_b^0) = 25.9\%$ \cite{LHCb:2019fns}, the production cross sections for $c\bar{c}$ and $b\bar{b}$ are measured to be 2369 $\mu b$ \cite{LHCb:2015swx} and 144 $\mu b$ \cite{LHCb:2016qpe} at the LHCb with $\sqrt{s} = 13 \ \rm TeV$. Supposing that LHCb will accumulate an integrated luminosity of approximately 50 $\rm fb^{-1}$ at the end of LHC Run-4 till 2035, hence we will get $N(\Lambda_c) = 2\times 50\ {\rm fb^{-1}} \times 2369 \ {\rm \mu b} \times 20.4\% = 4.83\times10^{13}$, $N(\Xi_c^0) = 2\times 50\ {\rm fb^{-1}} \times 2369 \ {\rm \mu b} \times 8\% = 1.90\times10^{13}$, $N(\Lambda_b) = 2\times 50\ {\rm fb^{-1}} \times 144 \ {\rm \mu b} \times 25.9\% = 3.73\times10^{12}$. The production cross section of $\Xi_c^+$ \cite{ALICE:2021bli} is measured to be 14.9 $\mu b$, the production number is estimated to be $N(\Xi_c^+) = 2\times 50\ {\rm fb^{-1}} \times 14.9 \ {\rm \mu b} = 1.49\times10^{12}$. The production cross section of $\Xi_{cc}^{++}$ can be estimated via its decay fraction versus $\Lambda_c^+$ \cite{LHCb:2019qed}, $\frac{\sigma(\Xi_{cc}^{++})Br(\Xi_{cc}^{++}\to\Lambda_c^+K^-\pi^+\pi^+)}{\sigma(\Lambda_c^+)} = 2.22\times10^{-4}$, and the fraction for $Br(\Xi_{cc}^{++}\to\Lambda_c^+K^-\pi^+\pi^+)$ is approximately 1.5\% \cite{Gutsche:2019iac}, hence the fraction is $\frac{\sigma(\Xi_{cc}^{++})}{\sigma(\Lambda_c^+)} = 1.48\%$, led to $N(\Xi_{cc}^{++}) = 4.83\times10^{13}\times1.48\% = 7.15\times10^{11}$. The production number of above heavy baryons are organized into TABLE.\ref{TabNBaryon}.

\begin{table}[ht]
	\begin{center}
		\caption{The reconstruction channels and detection efficiencies of heavy baryons $\Lambda_c,\ \Lambda_b,\ \Xi_{cc}^{++},\ \Xi_c^+,\ \Xi_c^0$ four-body decay. The branching fractions of secondary decay chain are adopted from \cite{ParticleDataGroup:2022pth}. The detection efficiency for $\Lambda_{b/c}$ is adopted according to estimation of previous researches \cite{Mejia-Guisao:2017nzx,Zhang:2021wjj}, the efficiencies for $\Xi_c^+,\ \Xi_c^0$ are estimated from similar decay channels at LHCb \cite{LHCb:2020gge}. Due to the experimental investigation of new discovered double-charmed $\Xi_{cc}^{++}$ is poor, hence we set the detection efficiency approximately to be $1\times10^{-4}$.}
		\label{TabChannelEfficiency}	
		\centering
		\begin{tabular}{|m{1.42cm}<{\centering}|m{2.1cm}<{\centering}|m{2.5cm}<{\centering}|m{2.5cm}<{\centering}|m{2.5cm}<{\centering}|m{2.5cm}<{\centering}|m{2.5cm}<{\centering}|}			
			\toprule
			\hline			
			& {\tiny $\Lambda_c^+\to\Lambda\mu^+\mu^+\pi^-$}  &  {\tiny $\Lambda_b^0\to\Lambda_c^+\mu^-\mu^-\pi^+$} & {\tiny $\Xi_{cc}^{++}\to\Xi_c^+\mu^+\mu^+\pi^-$} & {\tiny $\Xi_{cc}^{++}\to\Lambda_c^+\mu^+\mu^+\pi^-$} & {\tiny $\Xi_c^+\to\Xi^0\mu^+\mu^+\pi^-$} & {\tiny $\Xi_c^0\to\Xi^-\mu^+\mu^+\pi^-$}   \\
			\hline
			channel & {\tiny $\Lambda\to p\pi^- \ 64.1\%$}  & {\tiny $\Lambda_c^+\to pK^-\pi^+ \ 6.26\%$}          & {\tiny $\Xi_c^+\to pK^-\pi^+ \ 0.62\%$}           & {\tiny $\Lambda_c^+\to pK^-\pi^+ \ 6.26\%$}     & {\tiny $\Xi^0\to p\pi^-\pi^+ \ 63.8\%$} & {\tiny $\Xi^-\to p\pi^-\pi^- \ 64.0\%$}\\
			\hline
			$\epsilon_{eff}^{ful}$ & $1\times10^{-3}$\cite{Zhang:2021wjj} & $9.8\times10^{-3}$\cite{Mejia-Guisao:2017nzx} & $1\times10^{-4}$ & $1\times10^{-4}$ & $1.18\times10^{-2}$\cite{LHCb:2020gge} & $1.1\times10^{-3}$\cite{LHCb:2020gge}\\
			\hline
		\end{tabular}
	\end{center}		
\end{table}

In the fully reconstruction strategy, we explore the constraints of $|U_{\mu N}|^2$ from the experimental searches on $\Lambda_c^+\to\Lambda\mu^+\mu^+\pi^-, \Lambda_b^0\to\Lambda_c^+\mu^-\mu^-\pi^+, \Xi_{cc}^{++}\to\Xi_c^+\mu^+\mu^+\pi^-, \Xi_{cc}^{++}\to\Lambda_c^+\mu^+\mu^+\pi^-, \Xi_c^+\to\Xi^0\mu^+\mu^+\pi^-, \Xi_c^0\to\Xi^-\mu^+\mu^+\pi^-$ at the LHCb with the 95\% confidence level, that is $N(B_1)Br(B_1\to B_2+\mu\mu\pi)Br(B_2\to X)P_N\epsilon_{eff}^{ful} = 3.09$. Considering the values of the cross sections and efficiencies, the constraints for $|U_{\mu N}|^2$ versus $m_N$ at the 95\% confidence level are shown in FIG.\ref{FigB2Bllpi} at the LHCb in 13 TeV with an integrated luminosity of 50 $\rm fb^{-1}$. The final state baryons $\Lambda, \Xi^0, \Xi^-, \Lambda_c^+, \Xi_c^+$ are reconstructed via $\Lambda \to p\pi^-, \Xi^0 \to \Lambda(\to p\pi^-)\pi^+, \Xi^- \to \Lambda(\to p\pi^-)\pi^-, \Lambda_c \to pK^-\pi^+, \Xi_c^+ \to pK^-\pi^+$ channels respectively, the branching fraction are listed in TABLE \ref{TabChannelEfficiency}. The branching fraction for heavy baryon four-body decay reach its maximal value just above $\mu\pi$ threshold $\gtrsim 252 \rm MeV$, while the decay probabilities are highly suppressed by exponential factor, hence only moderate constraints for $|U_{\mu N}|^2$ are made. For $\Lambda_c \to \Lambda\mu^+\mu^+\pi^-$ channel, the maximal branching fraction can reach several part-per-thousand, and the $\Lambda_c$ can be numerously produced at LHCb, the lower limit for $|U_{\mu N}|^2$ can reach $10^{-4}\sim 10^{-5}$ at $0.25 < m_N < 1 \ {\rm GeV}$ through full reconstruction strategy. The fraction of charm quark fragment into $\Xi_c$ is smaller compare with $\Lambda_c^+$, and the branching fractions for $\Xi_c^+ \to \Xi^0\mu^+\mu^+\pi^-, \Xi_c^0 \to \Xi^-\mu^+\mu^+\pi^-$ are in part-per-thousand, hence the lower limits for $|U_{\mu N}|^2$ are suppressed by $1\sim 2$ magnitude at the same mass region, around $10^{-3}\sim 10^{-4}$. We also explore the decay channels for double-charmed baryon $\Xi_{cc}^{++}$ via $\Xi_{cc}^{++}\to\Xi_c^+\mu^+\mu^+\pi^-, \Xi_{cc}^{++}\to\Lambda_c^+\mu^+\mu^+\pi^-$, although the branching fractions are not small, the constraints are weak for less produced number and inefficiency. As pointed in previous research \cite{Zhang:2021wjj, Mejia-Guisao:2017nzx}, the bottom baryon $\Lambda_b^0$ can also provide unique test with broad mass region $0.3 \sim 3.0 \ \rm GeV$. As we can see, the lower limits for heavy baryon decay within full reconstruction method can only reach current bound around 2 GeV. 

In the partially reconstruction strategy, see FIG. \ref{FigB2Bllpi_V}, a distance of several millimeters, which is the typical flying length of B/D hadrons at LHC, is required from collision vertex to baryon decay vertex. Due to the fact that light and weak mixing sterile neutrino will fly long, the N decay vertex is generally far away from baryon decay vertex, e.g., in meters for $m_N =1 \ \rm GeV$ with $|U_{\ell N}|^2 = 10^{-4}$. To avoid background from B/D meson decay, at least one baryon is need in baryon decay vertex. In the "N Decay Vertex", similar decay signal due to $K\to\pi+\mu\nu$ can be avoided by requiring the direction of $\vec{p}_{\pi}+\vec{p}_{\mu}$ point to "Baryon Decay Vertex". At the 95\% confidence level, i.e., $N(B_1)Br(B_1\to B_2+\mu\mu\pi)P_N\epsilon_{eff}^{par} = 3.09$, where $\epsilon_{eff}^{par}$ is set to be $10\%$ due to the high efficiency to tag $\mu^{\pm}$ and $\pi^{\pm}$, the constraints for $|U_{\mu N}|^2$ is given in Fig. \ref{FigB2Bllpi}. As a full baryon event is avoided and the reconstruction efficiency is highly enhanced, the lower limits for $|U_{\mu N}|^2$ can be greatly  improved. For $\Lambda_b^0\to\Lambda_c^+\mu^-\mu^-\pi^+$ channel, a new bound around $10^{-5}\sim10^{-6}$ could be set for $2\sim3$ GeV. Considering a high luminosity and much yields of $\Lambda_b$ in High-Luminosity LHC (HL-LHC), the constraints will be further extended.

\subsection{Higgs Decay}
The collider searches for massive sterile neutrino via Higgs boson decay channels also provide a interesting test, especially for heavier mass as the Yukawa coupling is proportional to fermion mass. In the references \cite{BhupalDev:2012zg,Das:2017rsu,Das:2017zjc}, sterile neutrino is explored through direct coupling like ${\rm Higgs} \to \nu N$. Nevertheless, if the sterile neutrino mass lie much below Higgs mass, the coupling strength will dramatically suppressed, in this case the ${\rm Higgs} \to WW(\to \mu\mu\pi)$ channel may be also test window. Furthermore, in the direct Higgs decay channel one missing energy is required in the event construction; while in the ${\rm Higgs} \to WW(\to \mu\mu\pi)$ channel, final state products can be fully reconstructed and the $|\Delta L|=2$ same-sign $\mu\mu$ event exhibits absolutely new physics beyond SM.
\begin{figure}[ht]			
	\centering
	\subfigure{\includegraphics[scale=0.6]{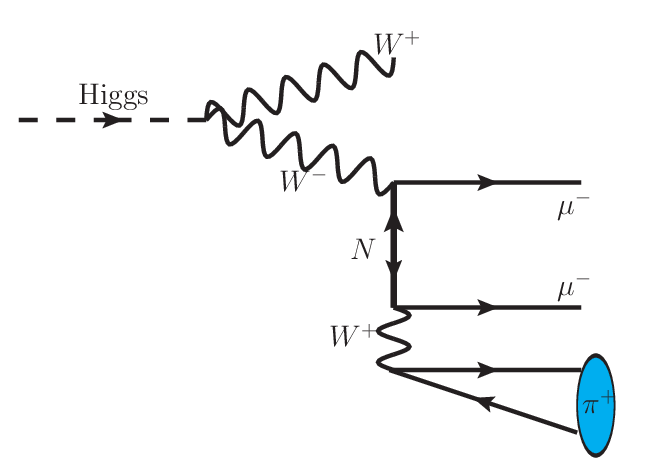}}
	\caption{Sterile neutrino production via ${\rm Higgs} \to WW(\to \mu\mu\pi)$ channel.}
	\label{FigH2WlN}
\end{figure}

In this section, we calculate the branching fraction for ${\rm Higgs} \to WW(\to \mu\mu\pi)$, the decay width can be formulated as
\begin{equation}
	\Gamma({\rm H} \to W^+\mu^-\mu^-\pi^+) = \Gamma({\rm H} \to W^+\mu^-N)\dfrac{\Gamma(N \to \mu^-\pi^+)}{\Gamma(N)},
\end{equation}
where narrow resonance approximation for sterile neutrino N is adopted. According to (\ref{EQLagrangian}), the decay amplitude is written as
\begin{equation}
	\mathcal{M}({\rm H} \to W^+\mu^-N) = -\dfrac{4m_W^3G_F U_{\mu N}}{(p_N+p_{\pi})^2-m_W^2}\epsilon^*_{\mu}(p_W)\bar{u}(p_\mu)\gamma^\mu\gamma_Lu(p_N),
\end{equation}
where $\epsilon^*_{\mu}(p_W)$ is the polarization vector of final $W^+$ boson, $\gamma_L = \dfrac{1-\gamma^5}{2}$. The amplitude square is
\begin{align}
	|\mathcal{M}({\rm H} \to W^+\mu^-N)|^2 &=
	\frac{16G_F^2m_W^4|U_{\mu N}|^2}{(m_H^2+m_{\mu}^2+m_N^2-s_1-s_3)^2}[m_H^2m_W^2+m_\mu^2(m_N^2+m_W^2-s_3) \nonumber\\
	&+(m_N^2+2m_W^2)(m_W^2-s_1)+s_3(s_1-2m_W^2)],
\end{align}
where we define $s_1 = (p_W+p_\mu)^2, s_2 = (p_W+p_N)^2$.

The decay width can be formulated as
\begin{align}
	\Gamma({\rm H} \to W^+\mu^-N) = \dfrac{1}{2m_H}\dfrac{1}{(2\pi)^5}\dfrac{\pi^2}{4m_H^2}\int_{(m_\mu+m_W)^2}^{(m_H-m_N)^2}ds_1\int_{s_2^{-}}^{s_2^{+}}ds_2 |\mathcal{M}({\rm H} \to W^+\mu^-N)|^2
\end{align}
Here $s_2^{\pm} = m_W^2+m_N^2-\frac{(s_1-m_H^2+m_N^2)(s_1+m_W^2-m_\mu^2)\mp\lambda^{1/2}(m_H^2,s_1,m_N^2)\lambda^{1/2}(s_1,m_W^2,m_\mu^2)}{2s_1}$, and $\dfrac{1}{(2\pi)^5}\dfrac{\pi^2}{4m_H^2}$ is the factor for three-body final state phase space. We set $m_H = 125\ \rm GeV, \Gamma_H = 3.2\ \rm MeV$ in the calculation.  

\begin{figure}[ht]			
	\centering
	\subfigure{\includegraphics[scale=0.27]{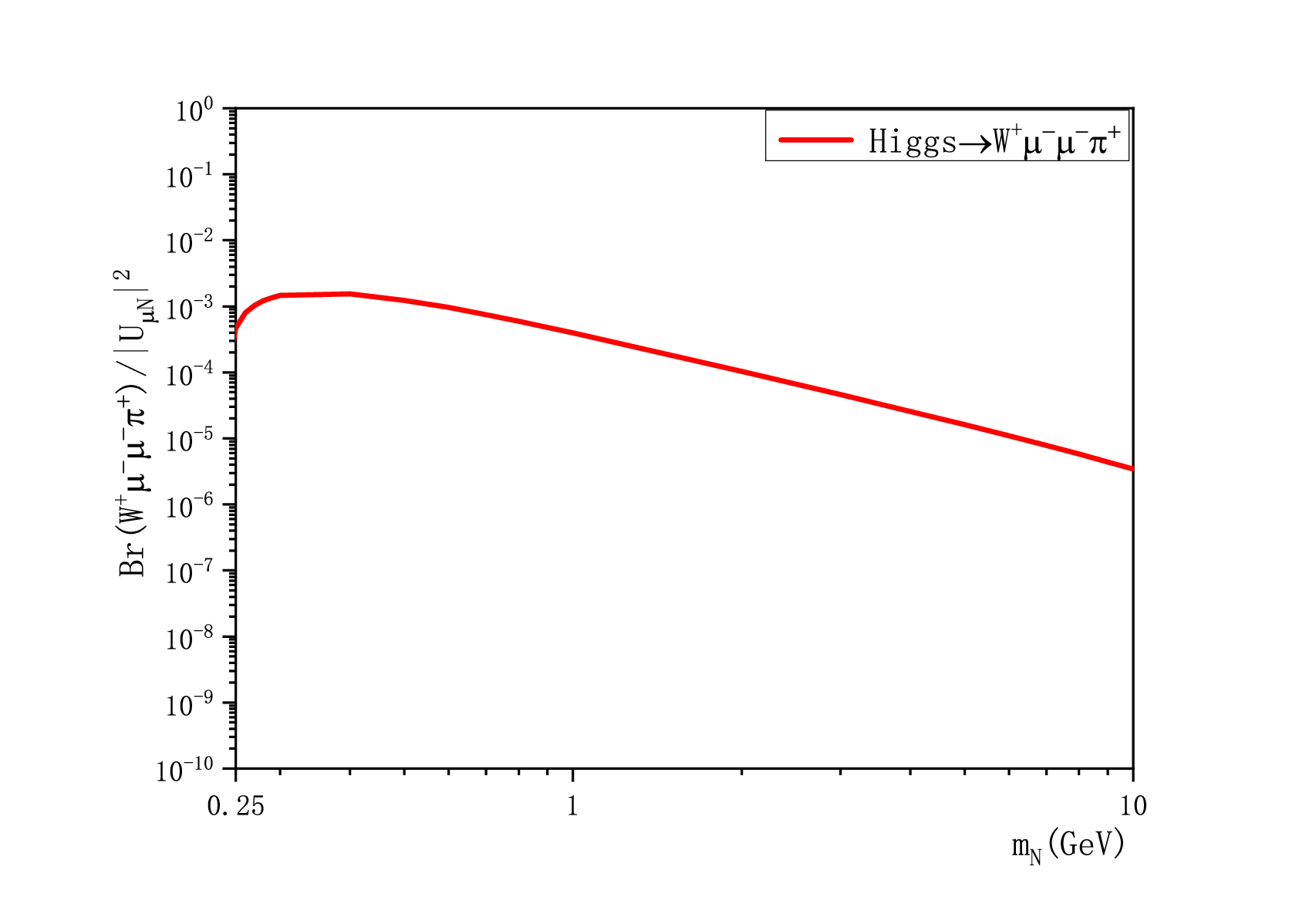}}
	\subfigure{\includegraphics[scale=0.27]{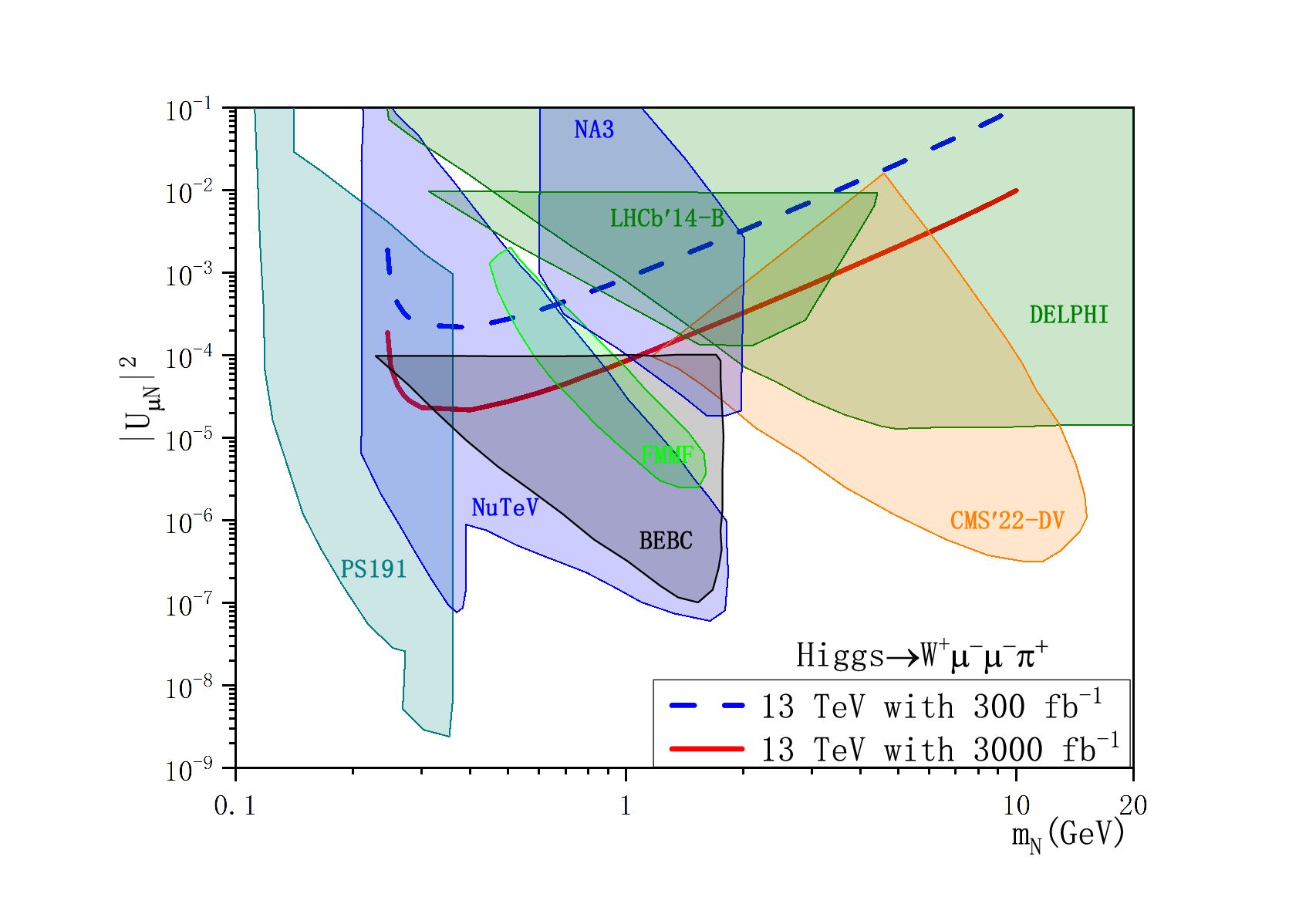}}	
	\caption{Branching fraction for ${\rm Higgs} \to WW(\to \mu\mu\pi)$ and the $|U_{\mu N}|^2$ sensitivity at the 95\% confidence level at 13 TeV LHC with integrate luminosity of 300/3000 $\rm fb^{-1}$ respectively. Here the production cross section of Higgs is estimated to be 50 pb, and the W boson is reconstructed via $c\bar{s}$-jet channel. The charge conjugate is hold here. The current experiment limits labeled ``PS191'' \cite{Bernardi:1987ek}, NA3 \cite{NA3:1986ahv}, BEBC \cite{WA66:1985mfx}, FMMF \cite{FMMF:1994yvb} are the constraints from beam-dump experiments, the label ``NuTeV'' \cite{NuTeV:1999kej} is the result from direct N decay search, the label ``LHCb$'$14-B'' is the constraints from B meson decay search, the constraint labeled ``DELPHI'' \cite{DELPHI:1996qcc} is given by $Z^0$ decay, and the labels ``CMS$'$22-DV'' \cite{CMS:2022fut} is the displaced vertex searches at the CMS detector.}
	\label{FigBrUlNHiggs}
\end{figure}

At the LHC, Higgs boson is mainly produced via gluon-gluon-fusion channel with cross section $\sigma_{ggF}(pp\to {\rm Higgs}+X) \sim$ 50 pb. The ATLAS has accumulated an integrated luminosity $\sim$ 150 $\rm fb^{-1}$, means $7.5\times10^6$ Higgs. After LHC update, the integrated luminosity will reach 3000 $\rm fb^{-1}$, hence $1.5\times10^{8}$ Higgs samples can be expected. Such abundant Higgs events will put new sight in sterile neutrino search. Compared with direct ${\rm Higgs} \to \nu N$ decay, the ${\rm Higgs} \to WW(\to \mu\mu\pi)$ channel can be easily detected and the new physics signal is much more clear. We analyze the branching fraction of this channel, see FIG.\ref{FigBrUlNHiggs}, with $m_N$ lie between $m_\mu+m_\pi$ and $m_H-m_W-m_\mu$. The fraction reaches maximal value at $m_N$ mass slightly above $m_\mu+m_\pi$ threshold, and decreases dramatically in high mass region. The $|U_{\mu N}|^2$ constraints versus $m_N$ at the 95\% confidence level are given in FIG.\ref{FigBrUlNHiggs},  as W boson is reconstructed via $c\bar{s}$-jet channel with $Br(W^+ \to c\bar{s}) = 30\%$. The lower-limit of $|U_{\mu N}|^2$, reach $10^{-5}$, lies just above $m_\mu+m_\pi$ threshold. Hence we recommend an experimental investigation in this mass region, the signal is obvious and the background is relatively clear.

\section{SUMMARY AND CONCLUSIONS}

\begin{figure}[htbp!]			
	\centering
	\subfigure{\includegraphics[scale=0.66]{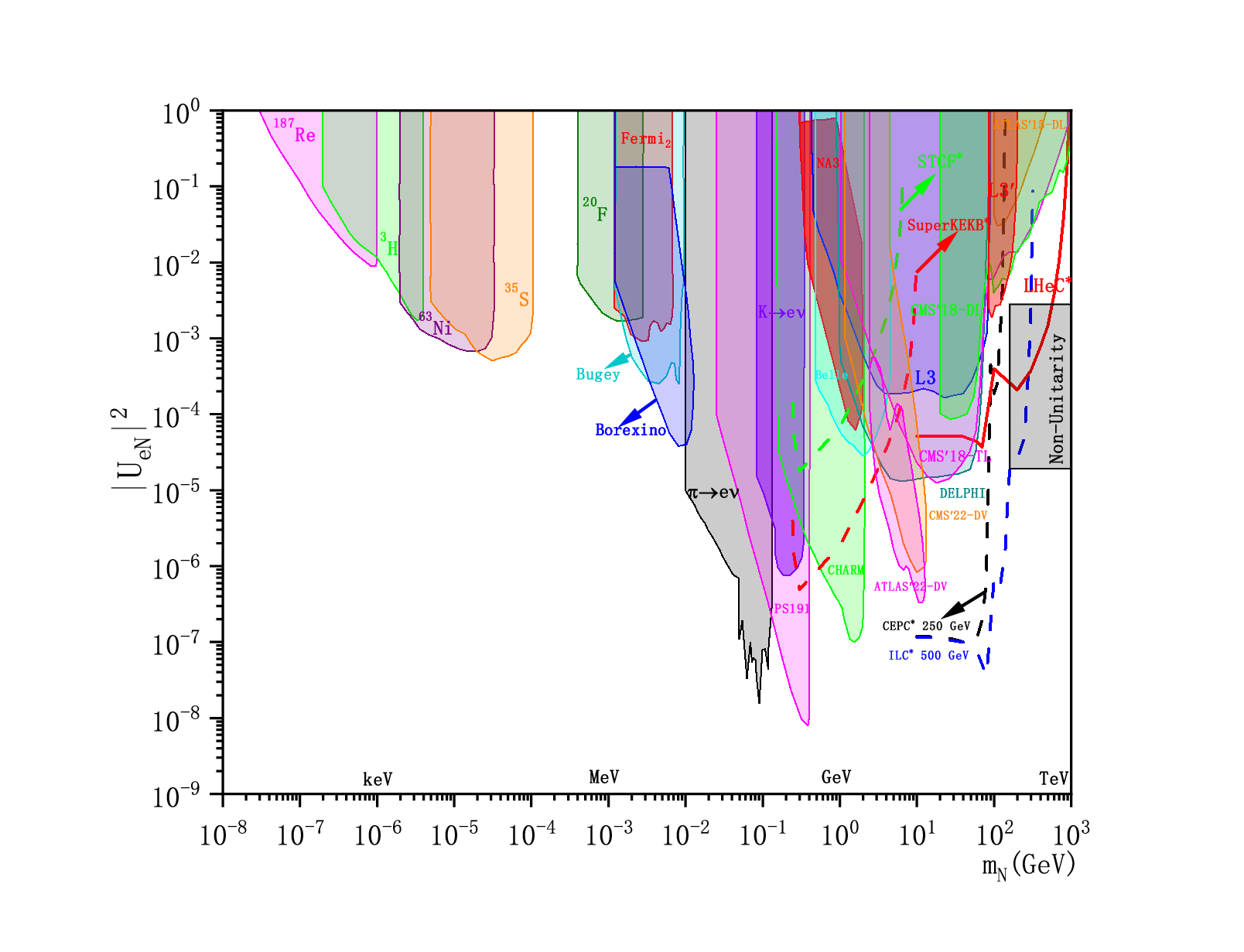}}	
	\caption{Bounds on active-sterile mixing $|U_{e N}|^2$ from various experiments and this article. Kink searches: $\rm ^{187}Re$ \cite{Galeazzi:2001py}, $\rm ^3H$ \cite{Hiddemann:1995ce}, $\rm ^{63}Ni$ \cite{Holzschuh:1999vy}, $\rm ^{35}S$ \cite{Holzschuh:2000nj}, $\rm ^{20}F$ \cite{Deutsch:1990ut}, $\rm Fermi_2$ \cite{Deutsch:1990ut}; N decay searches: Borexino \cite{Back:2003ae}, Bugey \cite{Hagner:1995bn}, L3$'$ \cite{L3:2001zfe}; Peak searches: $\pi\to e\nu$ \cite{Britton:1992pg}, $K\to e\nu$ \cite{Yamazaki:1984sj}; Beam-dump: PS191 \cite{Bernardi:1987ek}, NA3 \cite{NA3:1986ahv}, CHARM \cite{CHARM:1985nku}; $Z^0$ decay: DELPHI \cite{DELPHI:1996qcc}, L3 \cite{L3:1992xaz}; B meson decay: Belle \cite{Belle:2013ytx}; Direct production searches: ATLAS$'$15-DL \cite{ATLAS:2015gtp}, CMS$'$18-DL \cite{CMS:2018jxx}, CMS$'$18-TL \cite{CMS:2018iaf}; Displaced vertex searches: CMS$'$22-DV \cite{CMS:2022fut} and ATLAS$'$22-DV \cite{ATLAS:2022atq}; The label ``Non-Unitarity'' \cite{Blennow:2023mqx} is the constraints from non-unitarity effects. The curves labeled with ``*'' are the constraints from this paper.}
	\label{FigUeNExp}
\end{figure}

\begin{figure}[htbp!]			
	\centering
	\subfigure{\includegraphics[scale=0.66]{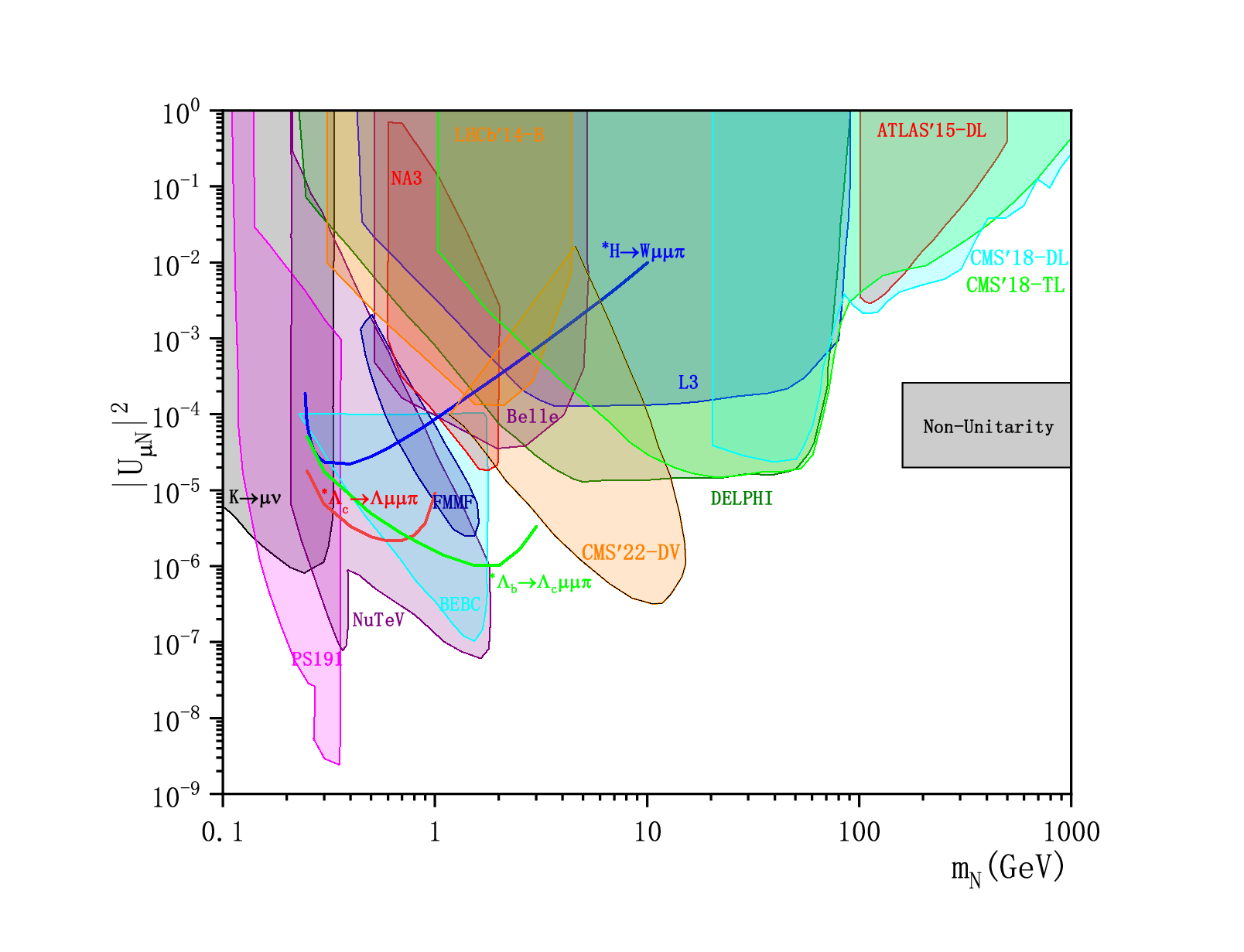}}	
	\caption{Current bounds on active-sterile mixing $|U_{\mu N}|^2$ from various experiments and this article. Peak searches: $K\to \mu\nu$ \cite{Kusenko:2004qc}; Beam-dump: PS191 \cite{Bernardi:1987ek}, NA3 \cite{NA3:1986ahv}, CHARMII \cite{CHARMII:1994jjr}, BEBC \cite{WA66:1985mfx}, FMMF \cite{FMMF:1994yvb}; $Z^0$ decay: DELPHI \cite{DELPHI:1996qcc}, L3 \cite{L3:1992xaz};  B meson decay: Belle \cite{Belle:2013ytx}, LHCb$'$14-B \cite{LHCb:2014osd}; N decay search: NuTeV \cite{NuTeV:1999kej}; Direct production searches: ATLAS$'$15-DL \cite{ATLAS:2015gtp}, CMS$'$18-DL \cite{CMS:2018jxx}, CMS$'$18-TL \cite{CMS:2018iaf}; Displaced vertex search: ATLAS$'$22-DV \cite{ATLAS:2022atq}. The label ``Non-Unitarity'' \cite{Blennow:2023mqx} is the constraints from non-unitarity effects. The curves labeled with ``*'' are the constraints from this paper.}
	\label{FigUmuNExp}
\end{figure}

In summary, we have studied the production mechanisms for sterile neutrino with collider signatures, the direct production channels at the $e^+e^-,\ ep$ colliders and indirect production through heavy particles decay are considered. Our results will provide complementary tests for the possible see-saw and radiative generation mass probes of sterile neutrino from GeV to TeV, the constraints for active-sterile mixing $|U_{\ell N}|^2$ will be extended. For the direct production channel at $e^+e^-$ collider, we investigate the W-exchange mechanism at the STCF, the SuperKEKB, the CEPC and the ILC, a novel signal cut method is proposed using open angle of $\ell jj$ and new limits on sterile neutrino mixing is expected. Numerical results indicate that the canonical cross section is considerable, the sensitivity for active-sterile mixing $|U_{\ell N}|^2$ is given at the 95\% confidence level. According to our estimation, the lower sensitivity limit can reach $10^{-3} \sim 10^{-6}$ in $0.3-2$ GeV region for lower energy $e^+e^-$ collider, $e.g.$, the SuperKEKB and the STCF; extended to $10^{-7}$ in electroweak energy mass region for high energy collider, $e.g.$, the CEPC and the ILC. For the direct production channel at $ep$ collider, we explore a new $\gamma$-$W^*$ fusion mechanism at the LHeC using proton bremsstrahlung. The canonical production cross section for $e^-\gamma\to NW^-$ process can reach tens fb if we probe a electroweak energy mass scale sterile neutrino, the low sensitivity limit of $|U_{eN}|^2$ will reach $10^{-4}$ on this mass region. Hundred of GeV heavy sterile neutrino can be also probe, new lower limit of sterile neutrino mixing on this mass region is expect. Besides, the Dirac or Majorana nature of the sterile neutrino can be tested via the sign of single lepton in $N\to\ell jj$ decay.

We also studied the indirect production channels via hadron (meson and baryon) and Higgs decay, better constraints on mixing than current experiment limits are given on related mass region. For heavy meson, we proposed a new search method for sterile neutrino via kink structure in lepton energy spectrum of B-meson semileptonic decay. For heavy baryon, we explore the four-body decay of $\Lambda_c, \Xi_{c}, \Xi_{cc}$ and $\Lambda_b$ under fully and partially reconstruction strategies. Considering the yields of heavy baryon is large at the LHCb,  the current $|U_{\mu N}|^2$ limits around $2\sim3\ \rm GeV$ can be extended to $10^{-5}\sim10^{-6}$. For Higgs decay, we investigate the $\rm Higgs\to W\mu\mu\pi$ process with relative clear signal, the constraint of mixing parameter is also given. The low sensitivity limits of $|U_{\ell N}|^2$ versus sterile neutrino mass from channels of this article and present experimental limits are given in FIG.\ref{FigUeNExp} - \ref{FigUmuNExp}.

\vspace{1.4cm} {\bf Acknowledgments}
The authors would like to thank Jingyi Xu for useful discussion on detection efficiency at the LHC. This work is supported in part by the National Key Research and Development Program of China under Contracts No. 2020YFA0406400, and by the National Natural Science Foundation of China (NSFC) under the Grants
Nos. 11975236, 12235008, 12275185 and 12335002.


\begin{thebibliography}{99}
	\bibitem{Minkowski:1977sc}
	P.~Minkowski,
	Phys. Lett. B \textbf{67}, 421-428 (1977)
	doi:10.1016/0370-2693(77)90435-X
	
	
	\bibitem{Mohapatra:1979ia}
	R.~N.~Mohapatra and G.~Senjanovic,
	Phys. Rev. Lett. \textbf{44}, 912 (1980)
	doi:10.1103/PhysRevLett.44.912
	
	
	\bibitem{Foot:1988aq}
	R.~Foot, H.~Lew, X.~G.~He and G.~C.~Joshi,
	Z. Phys. C \textbf{44}, 441 (1989)
	doi:10.1007/BF01415558
	
	
	\bibitem{Babu:1989fg}
	K.~S.~Babu and E.~Ma,
	Mod. Phys. Lett. A \textbf{4}, 1975 (1989)
	doi:10.1142/S0217732389002239
	
	
	\bibitem{Arkani-Hamed:1998wuz}
	N.~Arkani-Hamed, S.~Dimopoulos, G.~R.~Dvali and J.~March-Russell,
	Phys. Rev. D \textbf{65}, 024032 (2001)
	doi:10.1103/PhysRevD.65.024032
	[arXiv:hep-ph/9811448 [hep-ph]].
	
	
	\bibitem{Deppisch:2015qwa}
	F.~F.~Deppisch, P.~S.~Bhupal Dev and A.~Pilaftsis,
	New J. Phys. \textbf{17}, no.7, 075019 (2015)
	doi:10.1088/1367-2630/17/7/075019
	[arXiv:1502.06541 [hep-ph]].
	
	
	\bibitem{Doi:1985dx}
	M.~Doi, T.~Kotani and E.~Takasugi,
	Prog. Theor. Phys. Suppl. \textbf{83}, 1 (1985)
	doi:10.1143/PTPS.83.1
	
	
	\bibitem{Galeazzi:2001py}
	M.~Galeazzi, F.~Fontanelli, F.~Gatti and S.~Vitale,
	Phys. Rev. Lett. \textbf{86}, 1978-1981 (2001)
	doi:10.1103/PhysRevLett.86.1978
	
	
	\bibitem{Hiddemann:1995ce}
	K.~H.~Hiddemann, H.~Daniel and O.~Schwentker,
	J. Phys. G \textbf{21}, 639-650 (1995)
	doi:10.1088/0954-3899/21/5/008
	
	
	\bibitem{Holzschuh:1999vy}
	E.~Holzschuh, W.~Kundig, L.~Palermo, H.~Stussi and P.~Wenk,
	Phys. Lett. B \textbf{451}, 247-255 (1999)
	doi:10.1016/S0370-2693(99)00200-2
	
	
	\bibitem{Holzschuh:2000nj}
	E.~Holzschuh, L.~Palermo, H.~Stussi and P.~Wenk,
	Phys. Lett. B \textbf{482}, 1-9 (2000)
	doi:10.1016/S0370-2693(00)00476-7
	
	
	\bibitem{Deutsch:1990ut}
	J.~Deutsch, M.~Lebrun and R.~Prieels,
	Nucl. Phys. A \textbf{518}, 149-155 (1990)
	doi:10.1016/0375-9474(90)90541-S
	
	
	\bibitem{Britton:1992pg}
	D.~I.~Britton, S.~Ahmad, D.~A.~Bryman, R.~A.~Burnbam, E.~T.~H.~Clifford, P.~Kitching, Y.~Kuno, J.~A.~Macdonald, T.~Numao and A.~Olin, \textit{et al.}
	Phys. Rev. Lett. \textbf{68}, 3000-3003 (1992)
	doi:10.1103/PhysRevLett.68.3000
	
	
	\bibitem{Yamazaki:1984sj}
	T.~Yamazaki, T.~Ishikawa, Y.~Akiba, M.~Iwasaki, K.~H.~Tanaka, S.~Ohtake, H.~Tamura, M.~Nakajima, T.~Yamanaka and I.~Arai, \textit{et al.}
	Conf. Proc. C \textbf{840719}, 262 (1984)
	
	
	\bibitem{Bryman:2019ssi}
	D.~A.~Bryman and R.~Shrock,
	Phys. Rev. D \textbf{100}, no.5, 053006 (2019)
	doi:10.1103/PhysRevD.100.053006
	[arXiv:1904.06787 [hep-ph]].
	
	
	\bibitem{Bryman:2019bjg}
	D.~A.~Bryman and R.~Shrock,
	Phys. Rev. D \textbf{100}, 073011 (2019)
	doi:10.1103/PhysRevD.100.073011
	[arXiv:1909.11198 [hep-ph]].
	
	
	\bibitem{CLEO:2010ksb}
	P.~Rubin \textit{et al.} [CLEO],
	Phys. Rev. D \textbf{82}, 092007 (2010)
	doi:10.1103/PhysRevD.82.092007
	[arXiv:1009.1606 [hep-ex]].
	
	
	\bibitem{BELLE:2011bej}
	O.~Seon \textit{et al.} [BELLE],
	Phys. Rev. D \textbf{84}, 071106 (2011)
	doi:10.1103/PhysRevD.84.071106
	[arXiv:1107.0642 [hep-ex]].
	
	
	\bibitem{LHCb:2014osd}
	R.~Aaij \textit{et al.} [LHCb],
	Phys. Rev. Lett. \textbf{112}, no.13, 131802 (2014)
	doi:10.1103/PhysRevLett.112.131802
	[arXiv:1401.5361 [hep-ex]].
	
	
	\bibitem{BESIII:2019oef}
	M.~Ablikim \textit{et al.} [BESIII],
	Phys. Rev. D \textbf{99}, no.11, 112002 (2019)
	doi:10.1103/PhysRevD.99.112002
	[arXiv:1902.02450 [hep-ex]].
	
	
	\bibitem{L3:1992xaz}
	O.~Adriani \textit{et al.} [L3],
	Phys. Lett. B \textbf{295}, 371-382 (1992)
	doi:10.1016/0370-2693(92)91579-X
		
	
	\bibitem{DELPHI:1996qcc}
	P.~Abreu \textit{et al.} [DELPHI],
	Z. Phys. C \textbf{74}, 57-71 (1997)
	[erratum: Z. Phys. C \textbf{75}, 580 (1997)]
	doi:10.1007/s002880050370
	
	
	\bibitem{CMS:2022fut}
	A.~Tumasyan \textit{et al.} [CMS],
	JHEP \textbf{07}, 081 (2022)
	doi:10.1007/JHEP07(2022)081
	[arXiv:2201.05578 [hep-ex]].
	
	
	\bibitem{ATLAS:2022atq}
	G.~Aad \textit{et al.} [ATLAS],
	Phys. Rev. Lett. \textbf{131}, no.6, 061803 (2023)
	doi:10.1103/PhysRevLett.131.061803
	[arXiv:2204.11988 [hep-ex]].
	
	
	\bibitem{CMS:2012wqj}
	S.~Chatrchyan \textit{et al.} [CMS],
	Phys. Lett. B \textbf{717}, 109-128 (2012)
	doi:10.1016/j.physletb.2012.09.012
	[arXiv:1207.6079 [hep-ex]].
	
	
	\bibitem{CMS:2015qur}
	V.~Khachatryan \textit{et al.} [CMS],
	Phys. Lett. B \textbf{748}, 144-166 (2015)
	doi:10.1016/j.physletb.2015.06.070
	[arXiv:1501.05566 [hep-ex]].
	
	
	\bibitem{ATLAS:2015gtp}
	G.~Aad \textit{et al.} [ATLAS],
	JHEP \textbf{07}, 162 (2015)
	doi:10.1007/JHEP07(2015)162
	[arXiv:1506.06020 [hep-ex]].
	
	
	\bibitem{CMS:2018jxx}
	A.~M.~Sirunyan \textit{et al.} [CMS],
	JHEP \textbf{01}, 122 (2019)
	doi:10.1007/JHEP01(2019)122
	[arXiv:1806.10905 [hep-ex]].
	
	
	\bibitem{CMS:2018iaf}
	A.~M.~Sirunyan \textit{et al.} [CMS],
	Phys. Rev. Lett. \textbf{120}, no.22, 221801 (2018)
	doi:10.1103/PhysRevLett.120.221801
	[arXiv:1802.02965 [hep-ex]].
	
	
	\bibitem{L3:2001zfe}
	P.~Achard \textit{et al.} [L3],
	Phys. Lett. B \textbf{517}, 67-74 (2001)
	doi:10.1016/S0370-2693(01)00993-5
	[arXiv:hep-ex/0107014 [hep-ex]].
	
	
	\bibitem{ATLAS:2012ak}
	G.~Aad \textit{et al.} [ATLAS],
	Eur. Phys. J. C \textbf{72}, 2056 (2012)
	doi:10.1140/epjc/s10052-012-2056-4
	[arXiv:1203.5420 [hep-ex]].
	
	
	\bibitem{Blennow:2023mqx}
	M.~Blennow, E.~Fern\'andez-Mart\'\i{}nez, J.~Hern\'andez-Garc\'\i{}a, J.~L\'opez-Pav\'on, X.~Marcano and D.~Naredo-Tuero,
	JHEP \textbf{08}, 030 (2023)
	doi:10.1007/JHEP08(2023)030
	[arXiv:2306.01040 [hep-ph]].
					
	
	\bibitem{Pontecorvo:1957cp}
	B.~Pontecorvo,
	Sov. Phys. JETP \textbf{6}, 429 (1957)
	
	
	\bibitem{Maki:1962mu}
	Z.~Maki, M.~Nakagawa and S.~Sakata,
	Prog. Theor. Phys. \textbf{28}, 870-880 (1962)
	doi:10.1143/PTP.28.870
	
	
	\bibitem{Das:2012ze}
	A.~Das and N.~Okada,
	Phys. Rev. D \textbf{88}, 113001 (2013)
	doi:10.1103/PhysRevD.88.113001
	[arXiv:1207.3734 [hep-ph]].
	
	
	\bibitem{Das:2018usr}
	A.~Das, S.~Jana, S.~Mandal and S.~Nandi,
	Phys. Rev. D \textbf{99}, no.5, 055030 (2019)
	doi:10.1103/PhysRevD.99.055030
	[arXiv:1811.04291 [hep-ph]].
	
	
	\bibitem{Mekala:2022cmm}
	K.~Meka\l{}a, J.~Reuter and A.~F.~\.Zarnecki,
	JHEP \textbf{06}, 010 (2022)
	doi:10.1007/JHEP06(2022)010
	[arXiv:2202.06703 [hep-ph]].
	
	
	\bibitem{Das:2023tna}
	A.~Das, S.~Mandal and S.~Shil,
	Phys. Rev. D \textbf{108}, no.1, 015022 (2023)
	doi:10.1103/PhysRevD.108.015022
	[arXiv:2304.06298 [hep-ph]].
	
	
	\bibitem{Mekala:2023diu}
	K.~Meka\l{}a, J.~Reuter and A.~F.~\.Zarnecki,
	Phys. Lett. B \textbf{841}, 137945 (2023)
	doi:10.1016/j.physletb.2023.137945
	[arXiv:2301.02602 [hep-ph]].

  
    \bibitem{Banerjee:2015gca}
    S.~Banerjee, P.~S.~B.~Dev, A.~Ibarra, T.~Mandal and M.~Mitra,
    Phys. Rev. D \textbf{92}, 075002 (2015)
    doi:10.1103/PhysRevD.92.075002
    [arXiv:1503.05491 [hep-ph]].
     	
	
	\bibitem{Antusch:2016ejd}
	S.~Antusch, E.~Cazzato and O.~Fischer,
	Int. J. Mod. Phys. A \textbf{32}, no.14, 1750078 (2017)
	doi:10.1142/S0217751X17500786
	[arXiv:1612.02728 [hep-ph]].
	
	
	\bibitem{Abdullahi:2022jlv}
	A.~M.~Abdullahi, P.~B.~Alzas, B.~Batell, J.~Beacham, A.~Boyarsky, S.~Carbajal, A.~Chatterjee, J.~I.~Crespo-Anadon, F.~F.~Deppisch and A.~De Roeck, \textit{et al.}
	J. Phys. G \textbf{50}, no.2, 020501 (2023)
	doi:10.1088/1361-6471/ac98f9
	[arXiv:2203.08039 [hep-ph]].
	
	
	\bibitem{Hahn:2004fe}
	T.~Hahn,
	Comput. Phys. Commun. \textbf{168}, 78-95 (2005)
	doi:10.1016/j.cpc.2005.01.010
	[arXiv:hep-ph/0404043 [hep-ph]].
	
	
	\bibitem{Ding:2019tqq}
	J.~N.~Ding, Q.~Qin and F.~S.~Yu,
	Eur. Phys. J. C \textbf{79}, no.9, 766 (2019)
	doi:10.1140/epjc/s10052-019-7277-3
	[arXiv:1903.02570 [hep-ph]].
	
	
	\bibitem{Shen:2022ffi}
	Y.~F.~Shen, J.~N.~Ding and Q.~Qin,
	Eur. Phys. J. C \textbf{82}, no.5, 398 (2022)
	doi:10.1140/epjc/s10052-022-10301-4
	[arXiv:2201.05831 [hep-ph]].
	
	
	\bibitem{Blondel:2022qqo}
	A.~Blondel, C.~B.~Verhaaren, J.~Alimena, M.~Bauer, P.~Azzi, R.~Ruiz, M.~Neubert, O.~Mikulenko, M.~Ovchynnikov and M.~Drewes, \textit{et al.}
	Front. in Phys. \textbf{10}, 967881 (2022)
	doi:10.3389/fphy.2022.967881
	[arXiv:2203.05502 [hep-ex]].
	
	
	\bibitem{Atre:2009rg}
	A.~Atre, T.~Han, S.~Pascoli and B.~Zhang,
	JHEP \textbf{05}, 030 (2009)
	doi:10.1088/1126-6708/2009/05/030
	[arXiv:0901.3589 [hep-ph]].
	
	
	\bibitem{Zhang:2021wjj}
	G.~Zhang and B.~Q.~Ma,
	Phys. Rev. D \textbf{103}, no.3, 033004 (2021)
	doi:10.1103/PhysRevD.103.033004
	[arXiv:2101.05566 [hep-ph]].
	
	
	\bibitem{Helo:2010cw}
	J.~C.~Helo, S.~Kovalenko and I.~Schmidt,
	Nucl. Phys. B \textbf{853}, 80-104 (2011)
	doi:10.1016/j.nuclphysb.2011.07.020
	[arXiv:1005.1607 [hep-ph]].
	
	
	\bibitem{Milanes:2016rzr}
	D.~Milanes, N.~Quintero and C.~E.~Vera,
	Phys. Rev. D \textbf{93}, no.9, 094026 (2016)
	doi:10.1103/PhysRevD.93.094026
	[arXiv:1604.03177 [hep-ph]].
	
	
	\bibitem{Aeppli:1993cb}
	A.~Aeppli, F.~Cuypers and G.~J.~van Oldenborgh,
	Phys. Lett. B \textbf{314}, 413-420 (1993)
	doi:10.1016/0370-2693(93)91259-P
	[arXiv:hep-ph/9303236 [hep-ph]].
	
	
	\bibitem{Lindner:2016lxq}
	M.~Lindner, F.~S.~Queiroz, W.~Rodejohann and C.~E.~Yaguna,
	JHEP \textbf{06}, 140 (2016)
	doi:10.1007/JHEP06(2016)140
	[arXiv:1604.08596 [hep-ph]].
	
	
	\bibitem{Batell:2022ogj}
	B.~Batell, T.~Ghosh, T.~Han and K.~Xie,
	JHEP \textbf{03}, 020 (2023)
	doi:10.1007/JHEP03(2023)020
	[arXiv:2210.09287 [hep-ph]].
	
	
	\bibitem{Li:2018wut}
	S.~Y.~Li, Z.~G.~Si and X.~H.~Yang,
	Phys. Lett. B \textbf{795}, 49-55 (2019)
	doi:10.1016/j.physletb.2019.06.001
	[arXiv:1811.10313 [hep-ph]].
	
	
	\bibitem{Gluck:2002fi}
	M.~Gluck, C.~Pisano and E.~Reya,
	Phys. Lett. B \textbf{540}, 75-80 (2002)
	doi:10.1016/S0370-2693(02)02125-1
	[arXiv:hep-ph/0206126 [hep-ph]].
	
	
	\bibitem{Blaksley:2011ey}
	C.~Blaksley, M.~Blennow, F.~Bonnet, P.~Coloma and E.~Fernandez-Martinez,
	Nucl. Phys. B \textbf{852}, 353-365 (2011)
	doi:10.1016/j.nuclphysb.2011.06.021
	[arXiv:1105.0308 [hep-ph]].
	
	
	\bibitem{Antusch:2019eiz}
	S.~Antusch, O.~Fischer and A.~Hammad,
	JHEP \textbf{03}, 110 (2020)
	doi:10.1007/JHEP03(2020)110
	[arXiv:1908.02852 [hep-ph]].
	
	
	\bibitem{Shrock:1980vy}
	R.~E.~Shrock,
	Phys. Lett. B \textbf{96}, 159-164 (1980)
	doi:10.1016/0370-2693(80)90235-X
	
	
	\bibitem{Daum:1977ec}
	M.~Daum, G.~H.~Eaton, R.~Frosch, H.~Hirschmann, J.~McCulloch, R.~C.~Minehart and E.~Steiner,
	Phys. Lett. B \textbf{74}, 126-129 (1978)
	doi:10.1016/0370-2693(78)90077-1
	
	
	\bibitem{Belle-II:2018jsg}
	E.~Kou \textit{et al.} [Belle-II],
	PTEP \textbf{2019}, no.12, 123C01 (2019)
	[erratum: PTEP \textbf{2020}, no.2, 029201 (2020)]
	doi:10.1093/ptep/ptz106
	[arXiv:1808.10567 [hep-ex]].
	
	
	\bibitem{Ali:2001gsa}
	A.~Ali, A.~V.~Borisov and N.~B.~Zamorin,
	Eur. Phys. J. C \textbf{21}, 123-132 (2001)
	doi:10.1007/s100520100702
	[arXiv:hep-ph/0104123 [hep-ph]].
	
	
	\bibitem{Dib:2000wm}
	C.~Dib, V.~Gribanov, S.~Kovalenko and I.~Schmidt,
	Phys. Lett. B \textbf{493}, 82-87 (2000)
	doi:10.1016/S0370-2693(00)01134-5
	[arXiv:hep-ph/0006277 [hep-ph]].
	
	
	\bibitem{Shuve:2016muy}
	B.~Shuve and M.~E.~Peskin,
	Phys. Rev. D \textbf{94}, no.11, 113007 (2016)
	doi:10.1103/PhysRevD.94.113007
	[arXiv:1607.04258 [hep-ph]].
	
	
	\bibitem{Mandal:2017tab}
	S.~Mandal, M.~Mitra and N.~Sinha,
	Phys. Rev. D \textbf{96}, no.3, 035023 (2017)
	doi:10.1103/PhysRevD.96.035023
	[arXiv:1705.01932 [hep-ph]].
	
	
	\bibitem{Chun:2019nwi}
	E.~J.~Chun, A.~Das, S.~Mandal, M.~Mitra and N.~Sinha,
	Phys. Rev. D \textbf{100}, no.9, 095022 (2019)
	doi:10.1103/PhysRevD.100.095022
	[arXiv:1908.09562 [hep-ph]].
	
	
	\bibitem{ParticleDataGroup:2022pth}
	R.~L.~Workman \textit{et al.} [Particle Data Group],
	PTEP \textbf{2022}, 083C01 (2022)
	doi:10.1093/ptep/ptac097
	
	
	\bibitem{Belle:2019iji}
	M.~T.~Prim \textit{et al.} [Belle],
	Phys. Rev. D \textbf{101}, no.3, 032007 (2020)
	doi:10.1103/PhysRevD.101.032007
	[arXiv:1911.03186 [hep-ex]].
	
	
	\bibitem{Chang:1999gn}
	C.~H.~Chang, C.~D.~Lu, G.~L.~Wang and H.~S.~Zong,
	Phys. Rev. D \textbf{60}, 114013 (1999)
	doi:10.1103/PhysRevD.60.114013
	[arXiv:hep-ph/9904471 [hep-ph]].
	
	
	\bibitem{LHCb:2014set}
	R.~Aaij \textit{et al.} [LHCb],
	Int. J. Mod. Phys. A \textbf{30}, no.07, 1530022 (2015)
	doi:10.1142/S0217751X15300227
	[arXiv:1412.6352 [hep-ex]].
			
	
	\bibitem{Mejia-Guisao:2017nzx}
	J.~Mejia-Guisao, D.~Milanes, N.~Quintero and J.~D.~Ruiz-Alvarez,
	Phys. Rev. D \textbf{96}, no.1, 015039 (2017)
	doi:10.1103/PhysRevD.96.015039
	[arXiv:1705.10606 [hep-ph]].
	
	
	\bibitem{Shi:2019hbf}
	Y.~J.~Shi, W.~Wang and Z.~X.~Zhao,
	Eur. Phys. J. C \textbf{80}, no.6, 568 (2020)
	doi:10.1140/epjc/s10052-020-8096-2
	[arXiv:1902.01092 [hep-ph]].
	
	
	\bibitem{Shi:2019fph}
	Y.~J.~Shi, Y.~Xing and Z.~X.~Zhao,
	Eur. Phys. J. C \textbf{79}, no.6, 501 (2019)
	doi:10.1140/epjc/s10052-019-7014-y
	[arXiv:1903.03921 [hep-ph]].
	
	
	\bibitem{Azizi:2011mw}
	K.~Azizi, Y.~Sarac and H.~Sundu,
	Eur. Phys. J. A \textbf{48}, 2 (2012)
	doi:10.1140/epja/i2012-12002-1
	[arXiv:1107.5925 [hep-ph]].
	
	
	\bibitem{ALICE:2021dhb}
	S.~Acharya \textit{et al.} [ALICE],
	Phys. Rev. D \textbf{105}, no.1, L011103 (2022)
	doi:10.1103/PhysRevD.105.L011103
	[arXiv:2105.06335 [nucl-ex]].
	
	
	\bibitem{LHCb:2019fns}
	R.~Aaij \textit{et al.} [LHCb],
	Phys. Rev. D \textbf{100}, no.3, 031102 (2019)
	doi:10.1103/PhysRevD.100.031102
	[arXiv:1902.06794 [hep-ex]].
	
	
	\bibitem{LHCb:2015swx}
	R.~Aaij \textit{et al.} [LHCb],
	JHEP \textbf{03}, 159 (2016)
	[erratum: JHEP \textbf{09}, 013 (2016); erratum: JHEP \textbf{05}, 074 (2017)]
	doi:10.1007/JHEP03(2016)159
	[arXiv:1510.01707 [hep-ex]].
	
	
	\bibitem{LHCb:2016qpe}
	R.~Aaij \textit{et al.} [LHCb],
	Phys. Rev. Lett. \textbf{118}, no.5, 052002 (2017)
	[erratum: Phys. Rev. Lett. \textbf{119}, no.16, 169901 (2017)]
	doi:10.1103/PhysRevLett.118.052002
	[arXiv:1612.05140 [hep-ex]].
	
	
	\bibitem{ALICE:2021bli}
	S.~Acharya \textit{et al.} [ALICE],
	Phys. Rev. Lett. \textbf{127}, no.27, 272001 (2021)
	doi:10.1103/PhysRevLett.127.272001
	[arXiv:2105.05187 [nucl-ex]].
	
	
	\bibitem{LHCb:2019qed}
	R.~Aaij \textit{et al.} [LHCb],
	Chin. Phys. C \textbf{44}, no.2, 022001 (2020)
	doi:10.1088/1674-1137/44/2/022001
	[arXiv:1910.11316 [hep-ex]].
	
	
	\bibitem{Gutsche:2019iac}
	T.~Gutsche, M.~A.~Ivanov, J.~G.~K\"orner, V.~E.~Lyubovitskij and Z.~Tyulemissov,
	Phys. Rev. D \textbf{100}, no.11, 114037 (2019)
	doi:10.1103/PhysRevD.100.114037
	[arXiv:1911.10785 [hep-ph]].
	
	
	\bibitem{LHCb:2020gge}
	R.~Aaij \textit{et al.} [LHCb],
	Phys. Rev. D \textbf{102}, no.7, 071101 (2020)
	doi:10.1103/PhysRevD.102.071101
	[arXiv:2007.12096 [hep-ex]].
	
	
	\bibitem{BhupalDev:2012zg}
	P.~S.~Bhupal Dev, R.~Franceschini and R.~N.~Mohapatra,
	Phys. Rev. D \textbf{86}, 093010 (2012)
	doi:10.1103/PhysRevD.86.093010
	[arXiv:1207.2756 [hep-ph]].
	
	
	\bibitem{Das:2017rsu}
	A.~Das, Y.~Gao and T.~Kamon,
	Eur. Phys. J. C \textbf{79}, no.5, 424 (2019)
	doi:10.1140/epjc/s10052-019-6937-7
	[arXiv:1704.00881 [hep-ph]].
	
	
	\bibitem{Das:2017zjc}
	A.~Das, P.~S.~B.~Dev and C.~S.~Kim,
	Phys. Rev. D \textbf{95}, no.11, 115013 (2017)
	doi:10.1103/PhysRevD.95.115013
	[arXiv:1704.00880 [hep-ph]].
	
	
	\bibitem{Back:2003ae}
	H.~O.~Back, M.~Belata, A.~de Bari, T.~Beau, A.~de Bellefon, G.~Bellini, J.~Benziger, S.~Bonetti, C.~Buck and B.~Caccianiga, \textit{et al.}
	JETP Lett. \textbf{78}, 261-266 (2003)
	doi:10.1134/1.1625721
	
	
	\bibitem{Hagner:1995bn}
	C.~Hagner, M.~Altmann, F.~von Feilitzsch, L.~Oberauer, Y.~Declais and E.~Kajfasz,
	Phys. Rev. D \textbf{52}, 1343-1352 (1995)
	doi:10.1103/PhysRevD.52.1343
	
	
	\bibitem{Bernardi:1987ek}
	G.~Bernardi, G.~Carugno, J.~Chauveau, F.~Dicarlo, M.~Dris, J.~Dumarchez, M.~Ferro-Luzzi, J.~M.~Levy, D.~Lukas and J.~M.~Perreau, \textit{et al.}
	Phys. Lett. B \textbf{203}, 332-334 (1988)
	doi:10.1016/0370-2693(88)90563-1
	
	
	\bibitem{NA3:1986ahv}
	J.~Badier \textit{et al.} [NA3],
	Z. Phys. C \textbf{31}, 21 (1986)
	doi:10.1007/BF01559588
	
	
	\bibitem{CHARM:1985nku}
	F.~Bergsma \textit{et al.} [CHARM],
	Phys. Lett. B \textbf{166}, 473-478 (1986)
	doi:10.1016/0370-2693(86)91601-1
	
	
	\bibitem{Belle:2013ytx}
	D.~Liventsev \textit{et al.} [Belle],
	Phys. Rev. D \textbf{87}, no.7, 071102 (2013)
	[erratum: Phys. Rev. D \textbf{95}, no.9, 099903 (2017)]
	doi:10.1103/PhysRevD.87.071102
	[arXiv:1301.1105 [hep-ex]].
	
	
    \bibitem{Kusenko:2004qc}
    A.~Kusenko, S.~Pascoli and D.~Semikoz,
    JHEP \textbf{11}, 028 (2005)
    doi:10.1088/1126-6708/2005/11/028
    [arXiv:hep-ph/0405198 [hep-ph]].
	
	
	\bibitem{CHARMII:1994jjr}
	P.~Vilain \textit{et al.} [CHARM II],
	Phys. Lett. B \textbf{343}, 453-458 (1995)
	doi:10.1016/0370-2693(94)01422-9
	
	
	\bibitem{WA66:1985mfx}
	A.~M.~Cooper-Sarkar \textit{et al.} [WA66],
	Phys. Lett. B \textbf{160}, 207-211 (1985)
	doi:10.1016/0370-2693(85)91493-5
	
	
	\bibitem{FMMF:1994yvb}
	E.~Gallas \textit{et al.} [FMMF],
	Phys. Rev. D \textbf{52}, 6-14 (1995)
	doi:10.1103/PhysRevD.52.6
	
	
    \bibitem{NuTeV:1999kej}
    A.~Vaitaitis \textit{et al.} [NuTeV and E815],
    Phys. Rev. Lett. \textbf{83}, 4943-4946 (1999)
    doi:10.1103/PhysRevLett.83.4943
    [arXiv:hep-ex/9908011 [hep-ex]].
    \end{thebibliography}
\end{document}